\documentclass[aps,prd,reprint,superscriptaddress,nofootinbib,longbibliography,floatfix]{revtex4-2}

\usepackage[T1]{fontenc}
\usepackage[utf8]{inputenc}
\usepackage{amsmath,amssymb,bm,mathtools}
\usepackage{graphicx}
\usepackage{booktabs}
\usepackage{array}
\usepackage{xcolor}
\usepackage{hyperref}
\usepackage{microtype}
\usepackage{siunitx}
\usepackage{needspace}
\hypersetup{colorlinks=true,citecolor=blue,linkcolor=blue,urlcolor=blue}
\graphicspath{{figures/}}

\newcommand{\diag}{\operatorname{diag}}
\newcommand{\MeV}{\mathrm{MeV}}

\newcommand{\gtr}{g_{CS}^{\mathrm{tr}}}
\newcommand{\Cnf}{\mathcal C}
\newcommand{\Knf}{\mathcal K}

\begin{document}

\title{Mass and decay angles in the \texorpdfstring{$\Xi_c$--$\Xi_c'$}{Xi-c--Xi-c-prime} mixing tension: Virtual hybrids and a single-direction criterion}

\author{Sanghoon Kim}
\affiliation{Department of Computer Education, Gyeongkuk National University, Andong 36729, South Korea}

\author{Gihwan Nam}
\affiliation{Department of Physics, Yonsei University, Seoul 03722, South Korea}

\author{Jongheon Baek}
\email[Corresponding author: ]{bjh6941@yonsei.ac.kr}
\affiliation{Department of Physics and Institute of Physics and Applied Physics, Yonsei University, Seoul 03722, Korea}

\date{October 7, 2026}

\begin{abstract}
We compare three definitions of $\Xi_c$--$\Xi_c'$ mixing within a constituent-quark model: a spin-component probability angle, a common-source overlap angle, and an SU(3)-breaking mass-operator angle.
The probability angle is $3.11^\circ$, while the two operator-based definitions give $2.48$--$2.58^\circ$; a second parameter set gives $2.38$--$2.49^\circ$.
The operator-based angles are closer to the lattice determinations $1.2(1)^\circ$ and $0.810(50)(200)^\circ$, but remain larger.
We then couple conventional $qsQ$ states to virtual $qsQg$ configurations, where $q=u,d$ and $Q=c,b$.
An orbital-$S$-wave chromomagnetic transition containing both $j_{Qg}=1/2$ and $3/2$ raises the source and mass-operator charm angles to $2.67$--$3.16^\circ$ when normalized to reference short-range mass shifts.
The hybrid probability stays below $1.7\%$, and the corresponding bottom angles are $0.42$--$0.48^\circ$.
Eliminating the hybrid sector gives a rank-one mass correction precisely when all weighted transitions select the same retained-state combination.
The mode-resolved calculation separates the flavor-breaking response that changes the mixing angle from the response that generates a second transition direction.
The accompanying effective-current relation connects this mass analysis to radiative and semileptonic amplitudes through their current matrix elements.
\end{abstract}

\maketitle

\section{Introduction}

The $\Xi_c$ and $\Xi_c'$ baryons have the same quark content and total spin, but differ mainly in the spin of their light-quark pair~\cite{KornerKramerPirjol1994}.
Interactions that break flavor SU(3) and heavy-quark spin symmetry can mix these configurations.
The size of this mixing has been studied using mass relations, lattice correlation functions, and decay amplitudes~\cite{LiuEtAl2023,LiuWangZhang2024,Matsui2021,GengEtAl2023a}.
Each method probes the mixed states through a different operator.
We separate this dependence on the angle definition from the dynamics of the mixing.

The numerical results illustrate the problem.
A lattice two-point correlation-matrix analysis gives $\theta_c=1.2(1)^\circ$~\cite{LiuEtAl2023}.
A second method constructs flavor eigenstates at the SU(3)-symmetric point and evaluates the symmetry-breaking mass operator between them, giving $0.810(50)(200)^\circ$~\cite{LiuWangZhang2024}.
The estimated QED contribution is about $0.04^\circ$~\cite{DengEtAl2024QED}.
Larger values have been reported in continuum calculations: $5.5(1.8)^\circ$ from QCD sum rules~\cite{AlievEtAl2011} and $8.12(80)^\circ$ from heavy-quark effective theory~\cite{Matsui2021}.
Decay analyses have used angles near $16.27(2.30)^\circ$ and $24.66(90)^\circ$~\cite{KeFangShi2024,GengEtAl2023a,GengEtAl2023b}.
The latter value, $|\theta_c|=0.137(5)\pi$, reduces the semileptonic rate in that analysis by roughly twenty percent~\cite{GengEtAl2023a}.
A separate lattice form-factor calculation predicts ${\cal B}(\Xi_c^0\to\Xi^-e^+\nu_e)=3.58(12)\%$, above the measured values quoted in that study and consistent with approximate flavor symmetry~\cite{FarrellMeinel2025}.
The lattice angles are theoretical determinations, whereas the decay angles are inferred within the amplitude models used to analyze data.

We evaluate the probability, source-overlap, and mass-operator angles using the same constituent-quark wave functions.
The meson-calibrated parameters in Table~\ref{tab:parameters} define set A; a second valence parameter set, B, tests the sensitivity to the Hamiltonian.
The source and mass-operator constructions follow the lattice methods with nonrelativistic wave functions and operators.
For charm, they give angles near $2.5^\circ$ in set A and $2.4^\circ$ in set B, compared with a probability angle near $3.1^\circ$ in set A.

We next examine whether virtual hybrid components can reduce this difference.
Here a hybrid component means a color-singlet three-quark state with an explicit gluonic excitation, denoted $qqqg$.
The hybrid states enter as virtual intermediate configurations between conventional baryon states.
Eliminating them generates an energy-dependent correction to the conventional baryon Hamiltonian.
The same elimination also changes electromagnetic and weak current operators.
An apparent angle inferred from a decay amplitude can therefore differ from the angle that diagonalizes a mass matrix.
We derive both effective operators and calculate the resulting mass and wave-function corrections.

Hybrid baryons have been studied with flux-tube models, constituent-gluon models, and chromomagnetic lattice interpolators~\cite{CapstickPage1999,DudekEdwards2012,EdwardsEtAl2013,KhanRichardsWinter2020,CiminoWillemynsSemay2024,ViseurSemayChevalier2026,MeyerSwanson2015,BurkertEichmannKlempt2026}.
QCD sum rules have also been applied to charmed $qqqg$ states~\cite{YangEtAl2025}.
Most of these calculations determine hybrid-dominated masses.
We instead study their virtual contribution to conventional baryons.
The valence Hamiltonian, the fitted connected three-quark interaction, and the local $qqqg$ transition construction are taken from Ref.~\cite{BaekEtAl2026PPNP}.
Building on its model-space reduction, we resolve the intermediate modes and calculate their flavor, spin, and angle dependence.
We use the explicit hybrid self-energy as the short-distance correction and the fitted connected interaction only to set its reference strength.

We use $q=u,d$ for a light quark, $s$ for a strange quark, and $Q=c,b$ for the heavy quark.
For quark content $qsQ$ and total angular momentum and parity $J^P=1/2^+$, the two pair-spin functions are
\begin{equation}
 \begin{aligned}
 |\rho\rangle&=\left|[(qs)_{s_{qs}=0}Q]_{1/2}\right\rangle,\\
 |\lambda\rangle&=\left|[(qs)_{s_{qs}=1}Q]_{1/2}\right\rangle .
 \end{aligned}
 \label{eq:valence_spin_basis}
\end{equation}
Here $s_{qs}$ is the light-pair spin.
The conventional hyperfine interaction couples the two functions if $m_q\neq m_s$.
For a normalized valence wave function, let $P_\rho$ and $P_\lambda$ be its probabilities in the two spin components.
Writing $P_{\rm dom}=\max(P_\rho,P_\lambda)$ and $P_{\rm sub}=\min(P_\rho,P_\lambda)$, we define
\begin{equation}
 \theta_N=\tan^{-1}\sqrt{P_{\rm sub}/P_{\rm dom}}.
 \label{eq:norm_angle_definition}
\end{equation}
This probability-defined angle is nonnegative and insensitive to the relative phase of the components.
We distinguish it from the source-overlap angle $\theta_{\rm src}(R)$, where $R$ is the source radius, and the mass-operator angle $\theta_H$ defined in Sec.~\ref{sec:angle_definitions}.

A second question concerns the structure of the induced mass correction.
If every intermediate state couples to the same linear combination of the two retained states, the correction is proportional to a single rank-one projector.
Otherwise, it generally has rank two.
At a fixed energy, both cases admit diagonalization by a single rotation; the rank condition tests the more restrictive single-projector form of the correction.
We formulate the condition in Sec.~\ref{sec:formalism} and apply it to the constituent model in Sec.~\ref{sec:model}.
Section~\ref{sec:results} presents the numerical results and the current-operator analysis.
Section~\ref{sec:lattice} gives a finite-volume matching prescription, and Sec.~\ref{sec:limitations} states the limits of the calculation.

\section{Model-space reduction and the single-direction criterion}
\label{sec:formalism}

\subsection{Self-energy in the two-state space}

We divide the model space into conventional and hybrid sectors.
The projector $\mathsf P$ selects the two conventional baryon states retained in the calculation.
The projector $\mathsf Q$ selects the intermediate states to be eliminated, with $\mathsf P\mathsf Q=0$ and $\mathsf P+\mathsf Q=I$ within this model space.

Let $|v_\rho(x)\rangle$ and $|v_\lambda(x)\rangle$ be the two lowest eigenstates of the conventional Hamiltonian $H_{qqq}$.
They are labeled by their continuous connection to the spin functions in Eq.~\eqref{eq:valence_spin_basis} at the symmetric point $x=0$, where $x=(m_s-m_q)/(m_s+m_q)$.
Their spatial wave functions are included, and conventional $\rho$--$\lambda$ mixing is already present at nonzero $x$.
The retained block is $H_{\mathsf P\mathsf P}=\mathsf P H_{qqq}\mathsf P$.
The intermediate block $H_{\mathsf Q\mathsf Q}$ is the Hamiltonian acting within the resolved hybrid space; its orthonormal eigenstates and energies are $|h_n(x)\rangle$ and $E_{h_n}(x)$.

We separate one real, dimensionless transition strength $\gtr$ from the reduced transition operator $T$.
The operator $T_{\mathsf Q\mathsf P}\equiv T$ maps a retained state into the intermediate space, and $T_{\mathsf P\mathsf Q}=T^\dagger$ maps it back.
In each block subscript, the right index denotes the input space and the left index the output space.
The full Hamiltonian has the form
\begin{equation}
 H=\begin{pmatrix}
 H_{\mathsf P\mathsf P}&\gtr T^\dagger\\
 \gtr T&H_{\mathsf Q\mathsf Q}
 \end{pmatrix}.
 \label{eq:model_space_blocks}
\end{equation}
Both $T$ and its matrix representation
\begin{equation}
 W_{n\alpha}(x)=\langle h_n(x)|T|v_\alpha(x)\rangle,
 \quad \alpha=\rho,\lambda,
 \label{eq:transition_matrix_definition}
\end{equation}
use the reduced convention with the overall strength $\gtr$ factored out.
The matrix $W$ has one row for each intermediate mode and two columns for the retained states.
Its entries have dimensions of energy.

Let $E$ be the energy at which the effective Hamiltonian is evaluated.
Writing $|\psi_{\mathsf P}\rangle=\mathsf P|\Psi\rangle$ and $|\psi_{\mathsf Q}\rangle=\mathsf Q|\Psi\rangle$, the second block equation gives
\[
 |\psi_{\mathsf Q}\rangle=(E-H_{\mathsf Q\mathsf Q})^{-1}\gtr T|\psi_{\mathsf P}\rangle.
\]
Substitution into the first block equation gives
$[H_{\mathsf P\mathsf P}+\Sigma(E,x)]|\psi_{\mathsf P}\rangle=E|\psi_{\mathsf P}\rangle$, where
\begin{equation}
 \Sigma(E,x)=(\gtr)^2
 T_{\mathsf P\mathsf Q}(E-H_{\mathsf Q\mathsf Q})^{-1}T_{\mathsf Q\mathsf P}.
 \label{eq:feshbach_self_energy}
\end{equation}
The self-energy $\Sigma$ is the induced mass operator in the retained space.
This is the Feshbach reduction~\cite{Feshbach1958,StetcuRotureau2013}.

Below every retained intermediate level, $D_n(E,x)=E_{h_n}(x)-E$ is positive.
Let $\mathbf D=\operatorname{diag}(D_n)$.
We define the reduced positive matrix
\begin{equation}
 G(E,x)=W^\dagger(x)\mathbf D^{-1}(E,x)W(x).
 \label{eq:gram_intro}
\end{equation}
Its relation to the physical correction is
\begin{equation}
 \Delta H_{\rm hyb}(E,x)=\Sigma(E,x)=-(\gtr)^2G(E,x).
 \label{eq:induced_gram}
\end{equation}
The minus sign is fixed by the negative energy denominator $E-E_{h_n}=-D_n$.
Thus $G$ and $\Sigma$ describe the same eliminated-sector contribution, with the sign and common coupling displayed separately.
We denote the eigenvalues of $G$ by $0\leq\gamma_-\leq\gamma_+$.

The same matrix can be written without choosing hybrid eigenstates:
\begin{equation}
 G_{\alpha\beta}(E,x)
 =\langle v_\alpha|T^\dagger(H_{\mathsf Q\mathsf Q}-E)^{-1}T|v_\beta\rangle.
 \label{eq:operator_gram}
\end{equation}
Diagonalizing $H_{\mathsf Q\mathsf Q}$ converts this expression into Eq.~\eqref{eq:gram_intro}.
The matrix $G$ is invariant under basis changes inside the fixed intermediate space.
Its eigenvalue ratio $\gamma_-/\gamma_+$ is also invariant under unitary rotations of the retained states and independent of $\gtr$, provided $G\neq0$.

\paragraph*{Single-direction elimination criterion.}
Define the positive resolvent $\mathcal R_{\mathsf Q}(E)=(H_{\mathsf Q\mathsf Q}-E)^{-1}$.
For a nonzero transition operator,
\begin{equation}
 \begin{aligned}
 G&=T^\dagger\mathcal R_{\mathsf Q}T\\
  &=(\mathcal R_{\mathsf Q}^{1/2}T)^\dagger
    (\mathcal R_{\mathsf Q}^{1/2}T)
 \end{aligned}
 \label{eq:single_direction_criterion}
\end{equation}
has rank one if and only if the image of $\mathcal R_{\mathsf Q}^{1/2}T$ is one-dimensional.
Equivalently, all weighted transition rows select the same linear combination of $|v_\rho\rangle$ and $|v_\lambda\rangle$.
The induced interaction is then proportional to a single rank-one projector.
If two independent combinations are selected, $G$ has rank two.
A general real symmetric $2\times2$ Hamiltonian is diagonalized by one rotation.
The stronger single-projector condition is apparent in $G=\gamma_-I+(\gamma_+-\gamma_-)\Pi_+$, where $\Pi_+$ projects onto the larger-eigenvalue direction.
For rank two, the full correction also contains the identity term $\gamma_-I$; rank one corresponds to $\gamma_-=0$.
However, the transformed current contains its own matrix elements and can define a different angle, as discussed in Sec.~\ref{sec:full_mixing}.

The construction also applies to closed meson--baryon channels.
In this case, a channel sum and a momentum integral replace the discrete mode sum:
\begin{equation}
 G_{\alpha\beta}^{\rm loop}(E)=\sum_c\int dk\,
 \frac{\mathcal M_{c\alpha}^*(k)\mathcal M_{c\beta}(k)}{E_c(k)-E}.
 \label{eq:loop_gram}
\end{equation}
Here $c$ labels a channel, $k$ is its relative momentum, $E_c(k)$ its energy, and $\mathcal M_{c\alpha}(k)$ the reduced transition amplitude from state $\alpha$.
The integration measure is included in the amplitude convention.
For $E$ below all retained channel thresholds, the denominators are positive.
Writing $\bm m_c(k)=(\mathcal M_{c\rho}(k),\mathcal M_{c\lambda}(k))$, the continuous Cauchy--Binet identity gives
\begin{equation}
 \begin{aligned}
 \det G^{\rm loop}={}&\frac12\sum_{c,c'}\int dk\,dk'\\[-2pt]
 &\times\frac{|\bm m_c(k)\wedge\bm m_{c'}(k')|^2}
 {[E_c(k)-E][E_{c'}(k')-E]},
 \end{aligned}
 \label{eq:loop_cauchy_binet}
\end{equation}
where $\bm u\wedge\bm v=u_\rho v_\lambda-u_\lambda v_\rho$.
Even a single channel can give rank two if the ratio $\mathcal M_{c\rho}(k)/\mathcal M_{c\lambda}(k)$ depends on momentum.
The wedge-product expression in Eq.~\eqref{eq:loop_cauchy_binet} also covers zeros of either amplitude.
In a ${}^3P_0$ model, these amplitudes contain spin-recoupling factors and spatial overlap integrals~\cite{BarnesSwanson2008,BaekEtAl2026PPNP,CapstickRoberts2000}.
This gives a direct test for a common transition direction in an unquenched quark model.
The same construction can be used for $\Lambda_Q$--$\Sigma_Q$ isospin mixing with an $I=0,1$ retained basis and an isospin-breaking coordinate proportional to $m_d-m_u$.
For $\Xi_c$ mixing, possible channels with the required charm and strangeness include $\Lambda D$, $\Xi D_s$, and $\Sigma_c\bar K$; their partial waves must also have the required total angular momentum and parity.
Above an open threshold, the self-energy acquires an imaginary part and the pole problem requires a complex effective Hamiltonian~\cite{DoringEtAl2026}.

Equation~\eqref{eq:induced_gram} uses one reference energy for the retained pair.
For distinct unperturbed energies $E_\rho$ and $E_\lambda$, a Hermitian second-order Schrieffer--Wolff reduction instead uses~\cite{BravyiDiVincenzoLoss2011}
\begin{equation}
 \begin{aligned}
 (G_{\rm SW})_{\alpha\beta}={}&\frac12\sum_n W_{n\alpha}^*W_{n\beta}\\
 &\times\left[\frac{1}{E_{h_n}-E_\alpha}
              +\frac{1}{E_{h_n}-E_\beta}\right].
 \end{aligned}
 \label{eq:nondegenerate_sw}
\end{equation}
This symmetrized expression is Hermitian; positivity is not guaranteed for general transitions.
The rank criterion above applies to the positive common-energy Feshbach self-energy.
We compare the numerical entries obtained with the two prescriptions in the Supplemental Material.

\subsection{Cauchy--Binet identity and exact factorization}

The rank condition has a useful form in terms of individual modes.
Define the weighted transition matrix
\begin{equation}
 A=\mathbf D^{-1/2}W,
 \qquad G=A^\dagger A,
 \label{eq:weighted_transition}
\end{equation}
and let $w_n=(W_{n\rho},W_{n\lambda})$ be the $n$th row of $W$.
Because $A$ has two columns, the Cauchy--Binet identity gives
\begin{equation}
 \det G=
 \sum_{m<n}
 \frac{\left|\det\!\begin{pmatrix}w_m\\w_n\end{pmatrix}\right|^2}
 {D_mD_n}.
 \label{eq:cauchy_binet}
\end{equation}
Each term is nonnegative and compares the directions selected by two intermediate states.
For $G\neq0$, the determinant vanishes if and only if every two-row minor vanishes.
The matrix then has rank one.
For example, a transition matrix of the form
\begin{equation}
 w_n(x)=c_n(x)v^\dagger(x)
 \label{eq:factorized_tower}
\end{equation}
has $\gamma_-(x)=0$ for every $x$.
Here $c_n$ is a mode-dependent scalar and $v$ is a common two-component vector.
The radial modes can have different strengths, but they all select the same retained-state combination.
We call this the single-form-factor limit.

\begin{figure*}[t]
\includegraphics[width=0.92\textwidth]{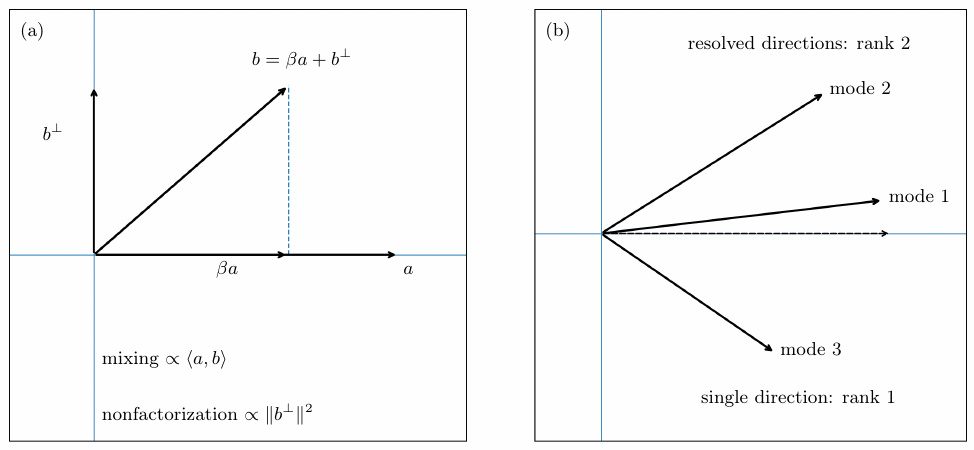}
\caption{Geometry of the single-direction condition.
Panel (a) separates the first flavor-breaking vector into $b=\beta a+b^\perp$.
The parallel part contributes to the off-diagonal mass matrix, while $\|b^\perp\|^2$ determines the leading smaller eigenvalue.
Panel (b) compares a common transition direction, which gives rank one, with mode-dependent directions, which can give rank two.}
\label{fig:single_direction_geometry}
\end{figure*}

For two intermediate modes, take
\begin{equation}
 A(x)=
 \begin{pmatrix}
 xb_1/\sqrt{D_1}&a_1/\sqrt{D_1}\\
 xb_2/\sqrt{D_2}&a_2/\sqrt{D_2}
 \end{pmatrix},
 \label{eq:two_mode_model}
\end{equation}
where $D_1,D_2>0$, at least one $a_i$ is nonzero, and the first column vanishes at $x=0$.
Expanding the smaller eigenvalue gives
\begin{equation}
 \gamma_-(x)=x^2
 \frac{|a_1b_2-a_2b_1|^2}
 {D_1D_2(|a_1|^2/D_1+|a_2|^2/D_2)}+O(x^4).
 \label{eq:two_mode_small_eigenvalue}
\end{equation}
The quadratic term vanishes if $a_1b_2-a_2b_1=0$.
If both $a_i$ are nonzero, this is the equal-slope condition $b_1/a_1=b_2/a_2$.
Thus two modes produce a second direction only if their relative response to flavor breaking differs.
Figure~\ref{fig:single_direction_geometry} illustrates the parallel and perpendicular components used below.

We now apply the expansion to each exchange-parity sector.
The operator that interchanges the light quarks $q$ and $s$ has eigenvalue $\eta=+1$ or $-1$ at $x=0$.
In the exchange-even sector, the transition from the $\rho$ state is forbidden at $x=0$, whereas that from the $\lambda$ state is allowed.
The weighted columns therefore have the form
\begin{equation}
 A_+(x)=\bigl[x\,b_++O(x^3),\ a_++O(x^2)\bigr].
 \label{eq:even_column_expansion}
\end{equation}
The columns are interchanged for the exchange-odd sector.
The vectors $a_\eta$ and $b_\eta$ have one component for each weighted hybrid mode: $a_\eta$ is the allowed transition at $x=0$, and $b_\eta$ is the linear response of the forbidden transition.
Assuming $a_\eta\neq0$, the two-column expansion gives
\begin{equation}
 \frac{\gamma_{\eta,-}}{\gamma_{\eta,+}}
 =x^2\Cnf_\eta^Q+O(x^4),
 \label{eq:ratio_expansion}
\end{equation}
where
\begin{align}
 \Cnf_\eta^Q
 =\frac{\|P_{a_\eta}^{\perp}b_\eta\|^2}{\|a_\eta\|^2},
 \qquad
 P_{a}^{\perp}
 =1-\frac{|a\rangle\langle a|}{\langle a|a\rangle}.
 \label{eq:nonfactorization_coefficient}
\end{align}
The operator $P_a^\perp$ removes the part of a vector parallel to $a$.
The coefficient $\Cnf_\eta^Q$ is nonnegative and vanishes if $b_\eta$ is proportional to $a_\eta$.
It measures the first flavor-breaking response that cannot be absorbed into a common retained-state rotation.
We use the term nonfactorization for this failure of a common transition direction.
Unitary basis changes within either sector leave $\Cnf_\eta^Q$ unchanged.
The corresponding dimensional coefficient is
\begin{equation}
 \Knf_\eta^Q=\lim_{x\to0}\frac{\gamma_{\eta,-}(x)}{x^2},
 \label{eq:dimensional_coefficient}
\end{equation}
and has units of MeV.
Multiplication by $(\gtr)^2$ gives the magnitude of the associated self-energy coefficient.

The intermediate space may contain orthogonal spin channels.
For channel $i$, decompose
\begin{equation}
 b_i=\beta_i a_i+b_i^\perp,
 \qquad
 n_i=\|a_i\|^2,
 \qquad
 \Knf_i=\|b_i^\perp\|^2,
 \label{eq:spin_channel_decomposition}
\end{equation}
with $\beta_i=\langle a_i|b_i\rangle/n_i$ and $n_i>0$.
For two channels, projection perpendicular to the combined allowed vector gives
\begin{equation}
 \Knf_{1\oplus2}
 =\Knf_1+\Knf_2+
 \frac{n_1n_2}{n_1+n_2}|\beta_1-\beta_2|^2 .
 \label{eq:spin_channel_mismatch_identity}
\end{equation}
The first two terms measure nonfactorization within each channel.
The last term measures the difference between their parallel slopes and can be nonzero even if each channel separately has rank one.
In contrast, the leading off-diagonal Gram element is
\begin{equation}
 G_{fa}(x)=x\sum_i\beta_i^* n_i+O(x^3),
 \label{eq:parallel_mixing_component}
\end{equation}
where $f$ and $a$ label the forbidden and allowed columns, respectively.
The complex conjugation follows from $G_{fa}=b^\dagger a$; all amplitudes and slopes in the numerical calculation below are real.
The off-diagonal mixing and $\Cnf_\eta^Q$ therefore test different components of the same response.
The former depends on the parallel part, and the latter on the perpendicular part.

This construction also separates the new transition response from the ordinary valence mixing.
If the forbidden retained state changes as
\begin{equation}
 |v_f(x)\rangle=|v_f(0)\rangle+x\,\theta_{\rm conv}'|v_a(0)\rangle+O(x^2),
 \label{eq:conventional_rotation}
\end{equation}
then its first transition vector changes by $b\to b+\theta_{\rm conv}'a$.
Here $|v_f(0)\rangle$ and $|v_a(0)\rangle$ are the forbidden and allowed valence states at the symmetric point.
Since
\begin{equation}
 P_a^\perp(b+\theta_{\rm conv}'a)=P_a^\perp b,
 \label{eq:conventional_rotation_cancellation}
\end{equation}
the leading coefficient $\Cnf_\eta^Q$ is unchanged by this conventional first-order rotation.
It isolates mode dependence that cannot be described by rotating both valence columns together.

\subsection{Constraints from light-flavor exchange}

Let $\mathcal X$ interchange $q$ and $s$, including their spatial and spin labels.
At equal light-quark masses, the retained spin states satisfy
\begin{equation}
 \mathcal X|\rho\rangle=-|\rho\rangle,
 \qquad
 \mathcal X|\lambda\rangle=+|\lambda\rangle,
 \label{eq:valence_exchange}
\end{equation}
and the hybrid states can be chosen with $\mathcal X|h_{\eta n}\rangle=\eta|h_{\eta n}\rangle$.
The common color phase of the retained singlet is factored out in this convention.
Interchanging the light masses sends $x$ to $-x$.
The forbidden transition column is therefore odd in $x$, while the allowed column is even.
Symmetry fixes the first permitted power in Eq.~\eqref{eq:ratio_expansion}, while the transition dynamics determine its coefficient.
If every linear two-row minor vanishes, the next permitted minor is cubic, so the ratio can begin at $x^6$.
It begins at still higher order if the cubic minors also vanish.

For the full matrix containing both exchange sectors,
\begin{equation}
 G(-x)=S G(x)S,
 \qquad S=\diag(-1,+1),
 \label{eq:full_gram_parity}
\end{equation}
up to continuous phase conventions.
The diagonal entries are even in $x$, and $G_{\rho\lambda}$ is odd.
These parity relations follow directly from exchange symmetry.
The values of $\Cnf_\eta^Q$, $\Knf_\eta^Q$, and their heavy-flavor ratios require the transition dynamics.
The small-eigenvalue expansion applies separately in each exchange sector; the full two-sector matrix can already have rank two at $x=0$.

\section{An orbital-\texorpdfstring{$S$-wave}{S-wave} constituent model}
\label{sec:model}

\subsection{Valence Hamiltonian and retained states}

We calculate the conventional baryon wave functions with the nonrelativistic Hamiltonian
\begin{equation}
 H_{qqq}=\sum_i\left(m_i+\frac{\bm p_i^2}{2m_i}\right)-K_{\rm CM}
 +\sum_{i<j}(V^C_{ij}+V^{CS}_{ij}),
 \label{eq:valence_hamiltonian}
\end{equation}
where $m_i$ and $\bm p_i$ are the constituent masses and momenta.
The center-of-mass kinetic energy is
$K_{\rm CM}=(\sum_i\bm p_i)^2/(2\sum_i m_i)$; subtracting it leaves the internal kinetic energy.
We set $c=1$ for masses and momenta and retain $\hbar c$ explicitly when lengths are in fm.
The central and color-spin potentials are
\begin{align}
 V^C_{ij}&=-\frac34F_i\!\cdot F_j
 \left(-\frac{\kappa}{r_{ij}}+\sigma r_{ij}-D\right),
 \label{eq:central_potential}\\
 V^{CS}_{ij}&=-\frac34F_i\!\cdot F_j
 \frac{\mu_{ij}\mu'_{ij}}{m_i m_j}
 \frac{\hbar c}{r_{ij}}e^{-(\mu_{ij}r_{ij}/\hbar c)^2}
 \bm\sigma_i\!\cdot\bm\sigma_j,
 \label{eq:hyperfine_potential}\\
 \mu_{ij}&=\alpha+\beta\frac{m_i m_j}{m_i+m_j},
 \qquad
 \mu'_{ij}=\gamma+\delta\frac{m_i m_j}{m_i+m_j}.
 \label{eq:range_parameters}
\end{align}
Here $r_{ij}=|\bm r_i-\bm r_j|$, $F_i^a=\lambda_i^a/2$ is a fundamental color generator, and $\bm\sigma_i$ is the Pauli-spin vector.
The parameters $\kappa$, $\sigma$, and $D$ determine the Coulomb term, confinement, and additive offset.
The scales $\mu_{ij}$ and $\mu'_{ij}$ determine the range and strength of the regulated hyperfine interaction.
The parameters in Table~\ref{tab:parameters} are the meson-calibrated short-range set of Ref.~\cite{BaekEtAl2026PPNP}.
The bottom calculation uses the same interaction parameters and $m_b$ from Table~\ref{tab:parameters}.

\begin{table}[t]
\caption{Constituent-model parameters for set A, taken from Ref.~\cite{BaekEtAl2026PPNP}.
Values are rounded for presentation.
The parameters $\beta$ and $\delta$ are dimensionless.}
\label{tab:parameters}
\begin{ruledtabular}
\footnotesize
\begin{tabular}{@{}lc@{\hspace{0.7em}}lc@{}}
Parameter & Value & Parameter & Value\\
\hline
$\kappa$ & $78.230~\MeV\,\mathrm{fm}$
& $\sigma$ & $1144.74~\MeV/\mathrm{fm}$\\
$D$ & $1098.22~\MeV$
& $\alpha$ & $285.198~\MeV$\\
$\beta$ & $0.666766$
& $\gamma$ & $167.289~\MeV$\\
$\delta$ & $0.150533$
& $m_q$ & $329.558~\MeV$\\
$m_s$ & $592.425~\MeV$
& $m_c$ & $1873.87~\MeV$\\
$m_b$ & $5298.0~\MeV$ & &
\end{tabular}
\end{ruledtabular}
\end{table}

The valence basis contains all three Jacobi rearrangements~\cite{HiyamaKinoKamimura2003}, with seven Gaussian $S$-wave ranges per coordinate between $0.05$ and $3.20~\mathrm{fm}$.
Both spin functions in Eq.~\eqref{eq:valence_spin_basis} are included in the same diagonalization.
The resulting eigenstates $|v_\rho(x)\rangle$ and $|v_\lambda(x)\rangle$ supply the two columns of transition amplitudes in $W$.
At the fitted light-quark masses, our lower-state norm angles are $3.114^\circ$ for charm and $0.895^\circ$ for bottom.
Equation~\eqref{eq:conventional_rotation_cancellation} explains why their leading rotation does not alter $\Cnf_\eta^Q$.

Table~\ref{tab:valence_spectrum} gives the valence masses before the hybrid correction, evaluated with the displayed parameters and the same seven-range basis.
The resulting splittings are $95.75~\MeV$ for charm and $127.69~\MeV$ for bottom.
The experimental columns use the neutral charm and negatively charged bottom states; the model is isospin symmetric and omits electromagnetic mass differences.
The residuals quantify the baryon-mass deviations for this fixed reference Hamiltonian.

\begin{table}[t]
\caption{Valence spectrum before the hybrid correction, in MeV.
The model values use Table~\ref{tab:parameters} and the seven-range Gaussian basis; experimental masses are from Ref.~\cite{ParticleDataGroup2024}.
The charge labels identify the experimental references for the isospin-symmetric model.}
\label{tab:valence_spectrum}
\begin{ruledtabular}
\begin{tabular}{l r r r}
State & $M_{qqq}$ & $M_{\rm exp}$ & $M_{qqq}-M_{\rm exp}$\\
\hline
$\Xi_c^0$ & $2489.64$ & $2470.44(28)$ & $+19.20$\\
$\Xi_c^{\prime0}$ & $2585.40$ & $2578.7(5)$ & $+6.70$\\
$\Xi_b^-$ & $5850.18$ & $5797.0(6)$ & $+53.18$\\
$\Xi_b^{\prime-}$ & $5977.87$ & $5935.1(5)$ & $+42.77$
\end{tabular}
\end{ruledtabular}
\end{table}

\subsection{Angle definitions and comparison with lattice calculations}
\label{sec:angle_definitions}

The norm angle $\theta_N$ measures spin-component probabilities in a variational wave function.
A lattice correlation-matrix angle follows from interpolator overlaps with the states~\cite{LiuEtAl2023}.
The SU(3)-breaking construction uses matrix elements of a mass operator between flavor eigenstates~\cite{LiuWangZhang2024}.
We construct the corresponding source-overlap and mass-operator angles from the same variational states.

For the source-overlap angle, define $\bm x_1=\bm r_q-\bm r_Q$ and $\bm x_2=\bm r_s-\bm r_Q$.
We use the same Gaussian profile for both flavor structures:
\begin{equation}
 \Phi_R\propto
 \exp\!\left[-\frac{x_1^2+x_2^2-\bm x_1\!\cdot\!\bm x_2}{3R^2}\right].
 \label{eq:gaussian_source}
\end{equation}
Here $R$ is the source radius.
Combining $\Phi_R$ with the flavor-$\overline{\bm3}$ and flavor-$\bm6$ spin structures defines source states $|O_a(R)\rangle$.
The overlaps with the two physical model states are $Z_{an}(R)=\langle O_a(R)|\Psi_n\rangle$, where $a=\overline{\bm3},\bm6$ and $n$ labels the lower and upper state.
We fit this matrix as
\begin{equation}
 Z(R)=\operatorname{diag}(z_{\overline3},z_6)
 \begin{pmatrix}
 \cos\theta_{\rm src}&\sin\theta_{\rm src}\\
 -\sin\theta_{\rm src}&\cos\theta_{\rm src}
 \end{pmatrix}
 +\delta Z .
 \label{eq:source_angle}
\end{equation}
The independent factors $z_{\overline3}$ and $z_6$ account for the normalization of each source.
The angle $\theta_{\rm src}$ describes the fitted rotation, and $\delta Z$ is the part not represented by that rotation.
The relative residual $\|\delta Z\|/\|Z\|$ is below $2.7\times10^{-5}$ for the cases quoted here.
The angle can still depend on $R$, because the two flavor components have different spatial wave functions.

\begin{figure*}[t]
\includegraphics[width=0.98\textwidth]{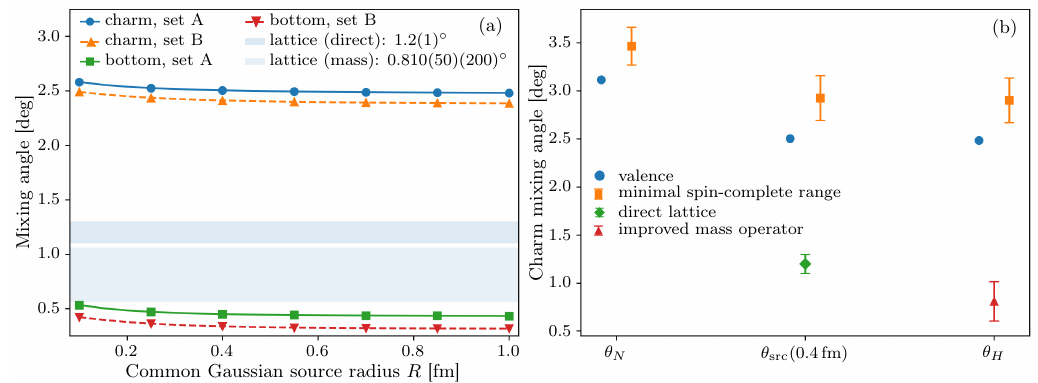}
\caption{Dependence on the angle definition.
Panel (a) shows the source-radius dependence for parameter sets A and B; the bands show the direct interpolator and improved mass-operator lattice results.
Panel (b) compares the set A charm values before and after the minimal spin-complete hybrid correction normalized to the reference short-range shifts.
The ranges vary the $j_{Qg}=1/2$ gap and the relative exchange-sector offset.
The two mass-operator lattice uncertainties are combined linearly in (a) ($0.250^\circ$) and in quadrature in (b) ($0.206^\circ$).
Including both spin sectors increases all three charm angles.}
\label{fig:angle_definitions}
\end{figure*}

For the mass-operator angle, define $m_0=(2m_q^{\rm fit}+m_s^{\rm fit})/3$ from the fitted masses.
We first set both light constituent masses to $m_0$ and construct the lowest flavor eigenstates $|a;m_0\rangle$ with energies $E_a^{(0)}$.
We then evaluate the physical two-state Hamiltonian:
\begin{equation}
 \begin{aligned}
 H^{(2)}_{ab}={}&E_a^{(0)}\delta_{ab}\\
 &+\langle a;m_0|H(m_q,m_s)-H(m_0,m_0)|b;m_0\rangle .
 \end{aligned}
 \label{eq:mass_operator_angle}
\end{equation}
The rotation that diagonalizes $H^{(2)}$ defines $\theta_H$.
This follows the sequence used in Ref.~\cite{LiuWangZhang2024}, but with a nonrelativistic energy operator and model wave functions.
The probability outside the two retained SU(3) states, called the two-state leakage, is $1.4$--$2.1\%$ in these calculations.
Figure~\ref{fig:angle_definitions} shows the angle comparisons; the detailed values and source-radius intervals are collected in Table~I and Sec.~II, respectively, of the Supplemental Material.

The norm angle is larger than the source and mass-operator angles for both heavy flavors, with a larger relative difference for bottom.
For charm with set A, the definition change lowers $3.11^\circ$ to $2.48$--$2.58^\circ$, a change of approximately $0.5$--$0.7^\circ$.
The charm mass-operator angles from sets A and B remain in the range $2.38$--$2.48^\circ$, above the improved lattice value $0.810(50)(200)^\circ$.
The source angles also remain above $1.2(1)^\circ$.

The small-$x$ rank test uses a different symmetric reference from the mass-operator construction.
Its trajectory keeps the average mass of the two light constituents fixed:
\begin{align}
 m_q(x)&=\bar m(1-x),
 &m_s(x)&=\bar m(1+x),
 \nonumber\\
 \bar m&=\frac{m_q^{\rm fit}+m_s^{\rm fit}}{2}.
 \label{eq:mass_trajectory}
\end{align}
The fitted masses give $x_{\rm phys}=0.28511$.
The coefficients $\Cnf_\eta^Q$ and $\Knf_\eta^Q$ are extracted near the symmetric point $x=0$.

\subsection{Hybrid color space and central Hamiltonian}

We retain the color-singlet component
\begin{equation}
 \left|[(qs)_{\overline{\mathbf 3}_C}(Qg)_{\mathbf 3_C}]_{\mathbf 1_C}\right\rangle,
 \label{eq:hybrid_color_state}
\end{equation}
with orbital $S$ waves and a gluonic excitation of spin and parity $j_g^P=1^+$.
The notation $j_{Qg}$ denotes the angular momentum obtained by coupling this excitation to the heavy-quark spin.
For $j_{Qg}=3/2$ and total $J=1/2$, the light pair has spin one.
The expectation values of the pair color operators are
\begin{equation}
\begin{array}{c|rrrrrr}
 ij & qs&qQ&sQ&qg&sg&Qg\\ \hline
 \langle F_i\cdot F_j\rangle
 &-2/3&1/12&1/12&-3/4&-3/4&-3/2
\end{array}
\label{eq:hybrid_color_factors}
\end{equation}
and sum to $-7/2$.
The full singlet color space also contains $[(qs)_{\mathbf 6_C}(Qg)_{\overline{\mathbf 6}_C}]_{\mathbf 1_C}$.
For the minimal emission map
$V|\xi\rangle=(\sqrt3/2)\sum_{a,\eta}(F^a)_{\eta\xi}|q_\eta g_a\rangle$, the final $Qg$ pair is in the unique $\mathbf3_C$ component of $\mathbf3_C\otimes\mathbf8_C$.
Here $\xi,\eta$ are quark color indices and $a$ is a gluon color index.
Projectors onto the other color representations give $P_{\overline{\mathbf6}}V=P_{\mathbf{15}}V=0$.
The minimal map therefore excites only the first singlet directly.
The calculation propagates this component; the second singlet requires color mixing in the hybrid Hamiltonian or additional transition operators.

The diagonal hybrid Hamiltonian contains internal kinetic energy and pairwise Cornell central potentials with the color factors in Eq.~\eqref{eq:hybrid_color_factors}.
The reference parameters are
\begin{equation}
 m_g=600~\MeV,\; r_g=0.30~\mathrm{fm},\; \Delta_{3/2}=870~\MeV,
 \label{eq:hybrid_benchmark_parameters}
\end{equation}
where $m_g$ is an effective gluon mass, $r_g$ is the width parameter of the Gaussian $Qg$ transition profile, and $\Delta_{3/2}$ sets the reference excitation gap of the $j_{Qg}=3/2$ sector.
Lattice and constituent-gluon studies motivate this gap scale~\cite{DudekEdwards2012,CiminoWillemynsSemay2024,ViseurSemayChevalier2026}; Sec.~\ref{sec:jhalf} examines the dependence on the spin-$1/2$ gap.

The four-body basis contains all 12 $K$-type and 6 ordered $H$-type Jacobi trees.
The $K$-type coordinates couple particles sequentially, while the $H$-type coordinates start from two pairs.
Each tree contains 15 Sobol-sampled triples of Gaussian ranges between $0.05$ and $3.0~\mathrm{fm}$, together with their light-flavor-exchanged partners, giving 540 basis functions before overlap reduction.
At $x=0$, removal of poorly conditioned overlap directions leaves 270 exchange-even and 210 exchange-odd vectors.
These subspaces are then kept at fixed dimension along the flavor-breaking trajectory.
This procedure avoids apparent rank changes caused only by basis vectors crossing an overlap cutoff.

\subsection{Chromomagnetic transition and heavy-quark dependence}

We use the local transition construction of Ref.~\cite{BaekEtAl2026PPNP}.
With the common coupling removed as in Eq.~\eqref{eq:model_space_blocks}, the reduced heavy-line vertex is
\begin{equation}
 T_{iQ}^{(3/2)}=
 \frac{1}{m_i m_Q}
 u_{iQ}(r_{iQ})
 \lambda_i^a(\lambda_Q^a)_{\rm tr}^\dagger
 \bm\sigma_i\cdot\bm{\mathsf S}_Q^\dagger,
 \qquad i=q,s.
 \label{eq:transition_operator}
\end{equation}
The physical off-diagonal Hamiltonian contains $\gtr T_{iQ}^{(3/2)}$.
The tensor $(\lambda_Q^a)_{\rm tr}^\dagger$ converts the heavy-quark color state into the retained color-triplet $Qg$ component.
The spin matrix $\bm{\mathsf S}_Q^\dagger$ connects the heavy spin-$1/2$ state to the $j_{Qg}=3/2$ state, while $\bm\sigma_i$ acts on the light quark.
The contraction is a rotational scalar.
Its inverse-mass factor has the same heavy-spin suppression as the hyperfine interaction.
The self-energy combines the explicit $1/m_Q^2$ factor with the heavy-mass dependence of the spatial overlaps and excitation denominators.

Projection onto Eq.~\eqref{eq:hybrid_color_state} gives the reduced color factor $-2/\sqrt3$.
In the initial $(\rho,\lambda)$ spin basis, the two light-quark paths give
\begin{equation}
 qQ:\ \left(\sqrt2,-\sqrt{\frac23}\right),
 \qquad
 sQ:\ \left(-\sqrt2,-\sqrt{\frac23}\right).
 \label{eq:j32_spin_rows}
\end{equation}
The radial function is
\begin{equation}
 u_{iQ}(r)=\mu_{iQ}\mu'_{iQ}\frac{\hbar c}{r}
 e^{-(\mu_{iQ}r/\hbar c)^2},
 \label{eq:transition_radial}
\end{equation}
together with a normalized Gaussian in the newly resolved $Qg$ coordinate.
It has dimensions $[u_{iQ}]=\MeV^3$, so the reduced transition matrix elements have units of MeV.
The operator and spatial profile specify the reduced transition, while $\gtr$ sets its overall normalization.
All reduced Gram matrices and dimensional coefficients are therefore quoted before multiplication by $(\gtr)^2$.

\section{Results for the virtual hybrid contribution}
\label{sec:results}

\subsection{Origin of the second transition direction}

We first study the $j_{Qg}=3/2$ sector and then add its spin-$1/2$ partner.
The exchange subspaces are defined at $x=0$ and kept at fixed dimension.
For each heavy flavor and exchange parity, we fit the small-$x$ behavior with
\begin{equation}
 y(x)=a_2x^2+a_4x^4,
 \qquad 0.001\leq x\leq0.008,
 \label{eq:constrained_coefficient_fit}
\end{equation}
and fix the intercept to zero.
Using $y=\gamma_-$ gives $\Knf_\eta^Q=a_2$.
The dimensionless coefficient is then
\begin{equation}
 \Cnf_\eta^Q=\frac{\Knf_\eta^Q}{\gamma_{\eta,+}(0)}.
 \label{eq:C_K_identity}
\end{equation}
A direct fit to $\gamma_-/\gamma_+$ provides an independent check.
The values reported below use Eq.~\eqref{eq:C_K_identity}.

To identify what produces a second direction, we start from a factorizing control calculation.
We solve the spin-independent central valence Hamiltonian at $x=0$ and use its ground spatial orbital $\phi_0$ for both spin configurations.
We keep the hybrid states and radial transition profile at that point and vary only
\begin{equation}
 f_i(x)=\frac{\mu_{iQ}(x)\mu'_{iQ}(x)}{m_i(x)m_Q}.
 \label{eq:scalar_vertex_factor}
\end{equation}
The light-flavor exchange symmetry then relates the two spatial overlaps by a sign, leaving one radial overlap $R_n$ for each mode.
The even and odd transition rows are
\begin{align}
 W_{+n}&\propto R_n
 \left(\sqrt2[f_q-f_s],-\sqrt{\frac23}[f_q+f_s]\right),
 \nonumber\\
 W_{-n}&\propto R_n
 \left(\sqrt2[f_q+f_s],-\sqrt{\frac23}[f_q-f_s]\right).
 \label{eq:physical_factorized_rows}
\end{align}
Within each exchange sector, the modes have the same retained-state direction.
All two-row minors vanish, and $\Cnf_\eta^Q=0$.
The common spatial orbital makes factorization exact even when the scalar mass factors vary.

We separate the effects of the orbitals, transition vertex, and state response through five approximations, summarized in Table~II of the Supplemental Material.
The common-orbital limit above provides the factorizing reference.
Keeping distinct symmetric-point orbitals for the two spin configurations, while varying only $f_i(x)$, isolates the effect of their spatial difference.
Varying the full radial vertex with the states fixed at $x=0$ gives the explicit vertex response.
Conversely, varying the states and excitation denominators with the vertex fixed at $x=0$ gives the state response.
Finally, holding the Gaussian transition range fixed while including the state response and scalar mass factors isolates the role of the range dependence.
The transition responses interfere, so their coefficients include cross terms.
Writing $P_a^\perp b_{\rm full}=u_{\rm vert}+u_{\rm resp}$ gives
\[
 \Cnf_{\rm full}=\Cnf_{\rm vert}+\Cnf_{\rm resp}
 +2\cos\phi\sqrt{\Cnf_{\rm vert}\Cnf_{\rm resp}},
\]
where $\cos\phi$ is the normalized real inner product of the two perpendicular response vectors.

A common orbital gives zero nonfactorization, and varying the mass factors alone gives only a small contribution.
The change of the radial vertex and the response of the states produce the larger nonparallel components.
The interference cosines are $+0.234,-0.673,+0.307,-0.636$ for $(c,+),(c,-),(b,+),(b,-)$, respectively.
The two mechanisms add constructively in the even sectors and partly cancel in the odd sectors.
The odd-sector coefficients are therefore more sensitive to changes in the spin-channel content.
Nonfactorization persists at fixed Gaussian width; the odd-sector coefficients are then larger than the full values.

\subsection{Charm-to-bottom ratios}

The coefficient $\Knf_\eta^Q$ contains the explicit heavy-spin factor, the mass-dependent radial normalization, spatial overlaps, and excitation denominators.
We separate these effects by repeating the calculation with all denominators set to $870~\MeV$.
Table~\ref{tab:hq_decomposition} expresses the charm-to-bottom ratio as a product of factors.

\begin{table}[t]
\caption{Multiplicative decomposition of $\Knf_\eta^c/\Knf_\eta^b$ in the set A $j_{Qg}=3/2$ model. The common reduced transition strength is assumed heavy-flavor independent.}
\label{tab:hq_decomposition}
\begin{ruledtabular}
\begin{tabular}{l c c}
Factor & $\eta=+$ & $\eta=-$ \\
\hline
naive $1/m_Q^2$ & $7.99$ & $7.99$ \\
$[\mu_c\mu_c'/(\mu_b\mu_b')]^2$ & $0.816$ & $0.816$ \\
spatial-overlap factor & $0.597$ & $0.429$ \\
hybrid-denominator factor & $0.949$ & $0.936$ \\
\hline
product & $3.70$ & $2.62$
\end{tabular}
\end{ruledtabular}
\end{table}

The radial normalization factor reduces the naive ratio from $7.99$ to $6.52$.
The spatial overlaps give the largest further reduction, while the different hybrid denominators have a smaller effect.
The resulting ratios are $3.70$ and $2.62$ in the even and odd sectors.
These model ratios assume the same reduced strength $\gtr$ for charm and bottom and depend on the retained spin channels, as examined next.

\subsection{Contribution of the \texorpdfstring{$j_{Qg}=1/2$}{jQg=1/2} sector}
\label{sec:jhalf}

Coupling a spin-one gluonic excitation to a spin-$1/2$ heavy quark also gives $j_{Qg}=1/2$.
The corresponding local transition matrices are $\mathsf R_\ell^\dagger=\sigma_{Q,\ell}/\sqrt3$, where $\ell=x,y,z$ labels a Cartesian component.
In the same phase convention as Eq.~\eqref{eq:j32_spin_rows}, the light-quark paths give
\begin{equation}
 M_q^{(1/2)}=
 \begin{pmatrix}0&-1\\-1&-2/\sqrt3\end{pmatrix},
 \qquad
 M_s^{(1/2)}=
 \begin{pmatrix}0&+1\\+1&-2/\sqrt3\end{pmatrix}.
 \label{eq:jhalf_recoupling}
\end{equation}
The rows label final light-pair spin $s_{qs}=0,1$, and the columns label initial $\rho,\lambda$ states.
The product of the spatial and light-pair spin parities gives the total exchange parity.

For the \mbox{spin-$3/2$} sector, $\mathsf S_\ell^\dagger=(\bm{\mathsf S}_Q^\dagger)_\ell$ is a Cartesian component of the spin-transition matrix appearing in Eq.~\eqref{eq:transition_operator}; its heavy-quark label $Q$ is suppressed.
The primed index $\ell'=x,y,z$ also labels a Cartesian component.
The matrix $\mathsf S_\ell^\dagger$ has dimensions $4\times2$ and maps the initial heavy-quark spin-$1/2$ space into the $j_{Qg}=3/2$ space; $\mathsf S_\ell$ is its adjoint.
The $2\times2$ matrices $\mathsf R_\ell^\dagger=\sigma_{Q,\ell}/\sqrt3$ introduced above instead map into the $j_{Qg}=1/2$ space.
The two products are
\[
 \begin{aligned}
 \mathsf S_\ell\mathsf S_{\ell'}^\dagger
 &=\delta_{\ell\ell'}I-\tfrac13\sigma_{Q,\ell}\sigma_{Q,\ell'},\\
 \mathsf R_\ell\mathsf R_{\ell'}^\dagger
 &=\tfrac13\sigma_{Q,\ell}\sigma_{Q,\ell'},
 \end{aligned}
\]
where $I$ is the identity in the two-dimensional heavy-quark spin space.
These products are the Cartesian blocks of the complementary projectors onto the two total-spin sectors of the spin-one times spin-$1/2$ product space, and therefore satisfy
\begin{equation}
 \mathsf S_\ell\mathsf S_{\ell'}^\dagger+
 \mathsf R_\ell\mathsf R_{\ell'}^\dagger=\delta_{\ell\ell'}I.
 \label{eq:spin_projector_completeness}
\end{equation}
For the minimal bare vector-creation operator, this fixes their relative reduced amplitude to $\zeta_{1/2}=1$.
Here $\zeta_{1/2}$ multiplies the spin-$1/2$ transition amplitude; a nonminimal value changes its self-energy contribution by $\zeta_{1/2}^2$.
With independent propagation of the orthogonal spin sectors, their Gram matrices add as $G=G_{3/2}+G_{1/2}$ at $\zeta_{1/2}=1$.
The relative phase cancels in this sum, whereas $\zeta_{1/2}\neq1$ introduces additional reduced transition structures.
We vary the excitation gaps to examine the diagonal $Qg$ spin dynamics independently of the transition vertex.
The spin-channel comparison in the Supplemental Material uses a common basis with 10 range triples per tree for the spin-$1/2$, \mbox{spin-$3/2$}, and combined results.

The large change in the odd sector follows from the first row of Eq.~\eqref{eq:jhalf_recoupling}.
A final light-pair spin-zero state does not couple locally to the initial $\rho$ state, and its two couplings to $\lambda$ have opposite signs.
For total exchange parity $-$, this channel has a spatially symmetric hybrid orbital.
Its $\lambda$ transition cancels at $x=0$, but flavor breaking produces a forbidden-column component without a parallel allowed-$\rho$ component.
It therefore contributes directly to $P_a^\perp b$ and can increase $\Cnf_-^Q$ by one to two orders of magnitude.
The analogous even-parity channel has a spatially antisymmetric orbital and a smaller contribution.

In this common basis, the dimensional charm-to-bottom ratios change from $3.48$ and $2.38$ for $j_{Qg}=3/2$ to $7.08$ and $6.96$ after spin completion.
The ratios are therefore sensitive to which spin channels are retained.
The Supplemental Material separates the intrinsic channel contributions from the slope-mismatch term in Eq.~\eqref{eq:spin_channel_mismatch_identity}.

The slope mismatch gives $78.1\%$ of the charm-even coefficient and $64.8\%$ of the bottom-even coefficient.
In the odd sectors, it gives only $3.3\%$ and $1.1\%$; the intrinsic spin-$1/2$ contribution dominates instead.
Thus the nonfactorization coefficient and the mixing-angle change probe the perpendicular and parallel responses, respectively, and can have different sensitivities to spin completion.
The Supplemental Material gives scans over $0\leq\zeta_{1/2}\leq2$ and a $\pm200~\MeV$ change of the spin-$1/2$ gap.

\subsection{Mass correction, effective currents, and \texorpdfstring{$\Xi_Q$}{XiQ} mixing}
\label{sec:full_mixing}

We now combine the two exchange sectors to obtain the correction to the conventional two-state Hamiltonian.
We first discuss the $j_{Qg}=3/2$ calculation, whose reduced hybrid space has 480 directions.
At $x=0$, the lowest even and odd levels have the same reference gap in the baseline model.
We keep the even reference fixed and vary the odd reference as
\begin{equation}
 \Delta_- = \Delta_{3/2}+\delta_-,
 \qquad -400\leq\delta_-\leq600~\MeV.
 \label{eq:sector_offset_definition}
\end{equation}
For equal offsets, the reduced Gram matrices at the fitted light-quark masses are
\begin{align}
 G_c&=
 \begin{pmatrix}3.754&-0.403\\-0.403&2.992\end{pmatrix}\MeV,
 \label{eq:full_charm_gram}\\
 G_b&=
 \begin{pmatrix}0.718&0.00959\\0.00959&0.5066\end{pmatrix}\MeV.
 \label{eq:full_bottom_gram}
\end{align}

\begin{table}[t]
\caption{Contributions to the off-diagonal element for the set A $j_{Qg}=3/2$ calculation at equal offsets.
Values are in MeV with $(\gtr)^2$ removed.
The cross contribution comes from the off-diagonal block of the hybrid resolvent after light-flavor breaking.}
\label{tab:g01_blocks}
\begin{ruledtabular}
\begin{tabular}{c r r r r}
$Q$&even block&odd block&cross block&total\\
\hline
$c$&$-0.7343$&$-0.2006$&$+0.5323$&$-0.4027$\\
$b$&$+0.1170$&$-0.00139$&$-0.1060$&$+0.00959$
\end{tabular}
\end{ruledtabular}
\end{table}

Table~\ref{tab:g01_blocks} separates the exchange-even, exchange-odd, and cross-block contributions.
The small off-diagonal element in Eq.~\eqref{eq:full_bottom_gram} reflects both heavy-quark suppression and a cancellation between sector contributions.
The exchange-even contribution is largely canceled by the cross-resolvent term, and the direct odd contribution is small.
Charm also has a substantial cancellation, but the odd contribution remains negative and keeps the total negative in this phase convention.
This decomposition describes the set A \mbox{spin-$3/2$} contribution.

For comparison with spin-dependent model interactions, we express the real symmetric two-state matrix using the effective spin-matrix convention of Appendix~\ref{app:spin}:
\begin{equation}
 \begin{aligned}
 G_Q(x)={}&\alpha_Q(x)I+\beta_Q(x)\,
 \bm\sigma_q\!\cdot\!\bm\sigma_s\\
 &+\chi_Q(x)\bigl(
 \bm\sigma_q\!\cdot\!\bm\sigma_Q-
 \bm\sigma_s\!\cdot\!\bm\sigma_Q\bigr).
 \end{aligned}
 \label{eq:matched_local_operator}
\end{equation}
The coefficients in this convention are
\begin{equation}
 \begin{aligned}
 \alpha_Q&=\frac{G_{\rho\rho}+3G_{\lambda\lambda}}{4},\\
 \beta_Q&=\frac{G_{\lambda\lambda}-G_{\rho\rho}}{4},
 \qquad\chi_Q=-\frac{G_{\rho\lambda}}{2\sqrt3}.
 \end{aligned}
 \label{eq:matched_operator_coefficients}
\end{equation}
The flavor-odd coefficient $\chi_Q$ represents the unequal light--heavy hyperfine structure in this two-state parametrization.
These coefficients parametrize $G_Q$ using the pure-spin matrices in Appendix~\ref{app:spin}.
Exchange symmetry makes $\alpha_Q$ and $\beta_Q$ even in $x$, and $\chi_Q$ odd.
Changing the relative exchange-sector centroid can move the off-diagonal entry for bottom through zero, while the corresponding entry for charm retains its sign over the same \mbox{spin-$3/2$} scan.

The same matrix can also be written in pairwise notation:
\begin{align}
 G_Q={}&A_QI+B_Q\,\bm\sigma_q\!\cdot\!\bm\sigma_s\nonumber\\
 &+C_{+,Q}(\bm\sigma_q\!\cdot\!\bm\sigma_Q+\bm\sigma_s\!\cdot\!\bm\sigma_Q)\nonumber\\
 &+C_{-,Q}(\bm\sigma_q\!\cdot\!\bm\sigma_Q-\bm\sigma_s\!\cdot\!\bm\sigma_Q).
 \label{eq:four_coefficient_operator}
\end{align}
There are only three independent real symmetric matrices in the retained two-state space.
The four coefficients above are therefore redundant, since
$\bm\sigma_q\!\cdot\!\bm\sigma_Q+\bm\sigma_s\!\cdot\!\bm\sigma_Q=-3I-\bm\sigma_q\!\cdot\!\bm\sigma_s$.
Choosing $A_Q=0$ gives
\begin{align}
 B_Q&=-\frac{G_{\rho\rho}}{3},&
 C_{+,Q}&=\frac{B_Q-G_{\lambda\lambda}}{4},\nonumber\\
 C_{-,Q}&=-\frac{G_{\rho\lambda}}{2\sqrt3}=\chi_Q.
 \label{eq:hyperfine_only_coefficients}
\end{align}
Table~\ref{tab:local_operator} gives the two equivalent parametrizations of the same calculated matrix.

\begin{table*}[t]
\caption{Physical-point projected operator coefficients for the set A $j_{Qg}=3/2$ tower. Values are in MeV before multiplication by $-(\gtr)^2$. The last three columns use the $A_Q=0$ convention of Eq.~\eqref{eq:hyperfine_only_coefficients}. Small-$x$ expansions are given in the Supplemental Material.}
\label{tab:local_operator}
\begin{ruledtabular}
\begin{tabular}{c r r r r r r}
$Q$&$\alpha_Q$&$\beta_Q$&$\chi_Q$&$B_Q$&$C_{+,Q}$&$C_{-,Q}$\\
\hline
$c$&$3.1822$&$-0.1905$&$+0.11625$&$-1.2512$&$-1.0607$&$+0.11625$\\
$b$&$0.55942$&$-0.05285$&$-0.002767$&$-0.23933$&$-0.18647$&$-0.002767$
\end{tabular}
\end{ruledtabular}
\end{table*}

The eliminated component also changes the state normalization.
For a normalized retained vector $p$, define
\begin{equation}
 \begin{aligned}
 \Omega_Q&=W^\dagger \mathbf D^{-2}W,\\
 P_{\rm hyb}&=\frac{(\gtr)^2p^\dagger \Omega_Qp}
 {1+(\gtr)^2p^\dagger \Omega_Qp}.
 \end{aligned}
 \label{eq:hybrid_probability}
\end{equation}
The dimensionless matrix $\Omega_Q$ determines the norm of the reconstructed hybrid component.
We obtain $p$ by diagonalizing the retained matrix $H_{\mathsf P\mathsf P}-(\gtr)^2G$ at the reference energy.
The numerator is the norm of its leading reconstructed hybrid component; the denominator normalizes the full state.

The same elimination must be applied to any external current~\cite{StetcuRotureau2013}.
For an external current $\mathcal J$, we define its blocks by
$\mathcal J_{\mathsf P\mathsf P}=\mathsf P\mathcal J\mathsf P$, $\mathcal J_{\mathsf P\mathsf Q}=\mathsf P\mathcal J\mathsf Q$, $\mathcal J_{\mathsf Q\mathsf P}=\mathsf Q\mathcal J\mathsf P$, and $\mathcal J_{\mathsf Q\mathsf Q}=\mathsf Q\mathcal J\mathsf Q$.
At the same reference energy, normalization of the reduced states gives
\begin{equation}
 \begin{aligned}
 \mathcal J_{\rm eff}={}&\mathcal J_{\mathsf P\mathsf P}-\gtr \mathcal J_{\mathsf P\mathsf Q}\mathbf D^{-1}W
 -\gtr W^\dagger \mathbf D^{-1}\mathcal J_{\mathsf Q\mathsf P}\\
 &+(\gtr)^2W^\dagger \mathbf D^{-1}\mathcal J_{\mathsf Q\mathsf Q}\mathbf D^{-1}W\\
 &-\frac{(\gtr)^2}{2}\{\Omega_Q,\mathcal J_{\mathsf P\mathsf P}\}+\cdots .
 \end{aligned}
 \label{eq:effective_current}
\end{equation}
The braces denote an anticommutator.
The linear terms are transition currents between the two sectors, the next term is the current acting inside the hybrid sector, and the last term comes from state normalization.
For a transition between different initial and final sectors, their own resolvents and normalization factors must be used; the corresponding form is given in the Supplemental Material.

The mass matrix $G$ and the effective current contain different matrix elements.
For a one-body impulse current that conserves explicit gluon number, $\mathcal J_{\mathsf P\mathsf Q}=\mathcal J_{\mathsf Q\mathsf P}=0$ and the leading correction is second order in the transition amplitude.
The charm probabilities below give a wave-function normalization scale of $0.3$--$1.6\%$.
For unsuppressed current matrix elements of comparable size in the two sectors, this suggests corrections at the percent level.
The relative correction also depends on the leading amplitude and can be enhanced by its suppression or by cancellations.
Transition currents instead enter linearly, with a wave-function scale $\sqrt{P_{\rm hyb}}\simeq5$--$13\%$ before their reduced current matrix elements are included.
Equation~\eqref{eq:effective_current} separates these contributions to radiative and semileptonic amplitudes.

The spectrum fit of Ref.~\cite{BaekEtAl2026PPNP} constrains the net connected interaction, which combines transition strengths, intermediate propagation, and potentially several unresolved contributions.
The fixed-parameter Yukawa shifts of the pure $\Xi_c$- and $\Xi_c'$-type pair-spin configurations are $-12.123$ and $-14.080~\MeV$.
Their magnitudes set the reference normalization of the hybrid contribution.
The full hybrid self-energy also contains induced one- and two-body terms, whereas the reference shifts describe the connected remainder.

At equal offsets, a least-squares fit of one minimal spin-complete strength to the two reference magnitudes gives $(\gtr)^2=1.268$.
The resulting diagonal magnitudes are $(8.24,16.07)~\MeV$, compared with $(12.123,14.080)~\MeV$.
Their ratio is $0.513$ rather than $0.861$.
A single strength fits the overall scale with residual differences in the two diagonal shifts.
The two reference shifts constrain the overall scale but leave the separate transition and propagation contributions undetermined.

The \mbox{spin-$3/2$} sector provides a diagnostic of its off-diagonal sign; the prediction for set A also includes the complementary spin-$1/2$ sector.
In unrestricted coupling scans, the lowest norm angles are $1.84^\circ$ for set A and $1.41^\circ$ for set B.
The source and mass-operator definitions can enter the lattice region in the spin-incomplete scan.

For the minimal vector vertex, $\zeta_{1/2}=1$, the one-strength reference normalization gives
\begin{align}
 3.268^\circ&\leq\theta_N(\Xi_c)\leq3.662^\circ,
 \nonumber\\
 0.279\%&\leq P_{\rm hyb}(\Xi_c)\leq1.606\%
 \label{eq:spin_complete_prediction_range}
\end{align}
over $670\leq\Delta_{1/2}\leq1070~\MeV$ and $-400\leq\delta_-\leq600~\MeV$.
The source and mass-operator angles are
\begin{align}
 2.693^\circ&\leq\theta_{\rm src}(0.4~\mathrm{fm})\leq3.158^\circ,
 \nonumber\\
 2.671^\circ&\leq\theta_H\leq3.135^\circ.
 \label{eq:spin_complete_lattice_like_range}
\end{align}
Their valence values are $2.504^\circ$ and $2.485^\circ$.
The spin-complete correction increases both charm angles and therefore widens their difference from the lattice values.

Using the same charm-normalized coupling for bottom gives
\begin{equation}
 \begin{aligned}
 0.423^\circ&\leq\theta_{\rm src}(0.4~\mathrm{fm})\leq0.454^\circ,\\
 0.450^\circ&\leq\theta_H\leq0.480^\circ,\\
 P_{\rm hyb}(\Xi_b)&<0.29\%.
 \end{aligned}
 \label{eq:xib_prediction_range}
\end{equation}
The valence values for set A are $0.449^\circ$ and $0.475^\circ$.
Set B gives a source interval $0.315$--$0.420^\circ$ and a mass-operator angle $0.359^\circ$.
The operator-based bottom angles are thus of order $0.3$--$0.5^\circ$, compared with the norm angle near $0.9^\circ$.
Table~\ref{tab:angle_probability_predictions} summarizes the set A valence and minimal spin-complete predictions for both heavy flavors.

\begin{table*}[t]
\caption{Set A angle and hybrid-probability results.
The spin-complete row scans the spin-$1/2$ gap and exchange-sector offset with the common strength normalized to the reference short-range shifts.
The last column combines the source and mass-operator bottom ranges; norm angles are listed separately.}
\label{tab:angle_probability_predictions}
\begin{ruledtabular}
\begin{tabular}{l c c c c c}
case&$\theta_{N,c}$&$\theta_{{\rm src},c}(0.4~\mathrm{fm})$&$\theta_{H,c}$&$P_{\rm hyb}^c$&bottom lattice-like angle\\
\hline
set A valence&$3.114^\circ$&$2.504^\circ$&$2.485^\circ$&$0$&$0.449/0.475^\circ$\\
minimal spin complete&$3.268$--$3.662^\circ$&$2.693$--$3.158^\circ$&$2.671$--$3.135^\circ$&$0.279$--$1.606\%$&$0.423$--$0.480^\circ$
\end{tabular}
\end{ruledtabular}
\end{table*}

A wider scan over $0\leq\zeta_{1/2}\leq2$ gives $-1.25^\circ\leq\Delta\theta_N(\Xi_c)\leq+1.03^\circ$, with $P_{\rm hyb}<1.7\%$.
Section XIII of the Supplemental Material gives a unit-coupling benchmark of the set A \mbox{spin-$3/2$} limit using the same retained wave functions as the offset calculation.
This wider range reflects changes to the transition operator, while Eq.~\eqref{eq:spin_complete_lattice_like_range} keeps the minimal vertex fixed.

For a small correction, the signed rotation of the retained eigenbasis is
\begin{equation}
 \delta\varphi_{\rm 2st}
 =\frac{(\gtr)^2G_{\rho\lambda}}{E_\lambda-E_\rho},
 \label{eq:two_state_rotation}
\end{equation}
where $E_\rho,E_\lambda$ are the unperturbed valence energies.
This first-order eigenbasis rotation and the probability-angle change differ because the spin components contain different spatial amplitudes.
At equal offsets and unit \mbox{spin-$3/2$} strength, exact diagonalization gives a retained-basis rotation of $-0.239^\circ$, while the norm-angle change is $-0.189^\circ$.
The model splitting is $95.75~\MeV$, about $11\%$ below the observed $\Xi_c'-\Xi_c$ separation~\cite{ParticleDataGroup2024}.
Using the observed splitting in the linear formula reduces its magnitude by about the same fraction.

\subsection{Numerical and model dependence}

The small-$x$ calculation keeps the 270 even and 210 odd overlap directions selected at $x=0$.
With the fit in Eq.~\eqref{eq:constrained_coefficient_fit}, the coefficient obtained directly from $\gamma_-/\gamma_+$ agrees with $\Knf_\eta^Q/\gamma_{\eta,+}(0)$ to within a relative difference of $10^{-4}$ in each sector.

We assess basis convergence with an independent nested Sobol sequence for each Jacobi tree, retaining the existing vectors as the basis grows to 30 range triples per tree.
The largest $25\to30$ change is $5.3\%$.
An independent 30-triple geometric-log basis agrees within $0.1$--$5.0\%$, and the 15-triple coefficients differ from the nested 30-triple values by $0.4$--$2.2\%$.
These comparisons give a $6\%$ basis uncertainty for the integrated coefficients.
Section~XVII of the Supplemental Material gives the full convergence tests.

Changes of $m_g$, $r_g$, and the diagonal central interaction alter the coefficients but preserve the leading power allowed by exchange symmetry.
For $0.25\leq r_g\leq0.40~\mathrm{fm}$, the all-mode $\Cnf_\eta^Q$ changes by at most $5.7\%$.
Changing the reference gap from $700$ to $1050~\MeV$ changes it by at most $9.7\%$.
Replacing the Gaussian radial transition by a Yukawa form with the same three-dimensional integral and root-mean-square (rms) radius changes $\Cnf_\eta^Q$ by $8.8$--$17.1\%$.
Here the rms radius of $u_{iQ}(r)$ in Eq.~\eqref{eq:transition_radial} is defined by
\[
 r_{{\rm rms},u}^2
 =\frac{\int d^3r\,r^2u_{iQ}(r)}{\int d^3r\,u_{iQ}(r)}
 =\left(\frac{\hbar c}{\mu_{iQ}}\right)^2.
\]
The moment uses the transition profile $u_{iQ}$ itself as the weight.
The explicit radial replacement is given in Sec.~XV of the Supplemental Material.
Both radial forms give nonzero $\Cnf_\eta^Q$ in all four sectors.

Parameter set B changes the charm-even $\Knf_+^c$ from $0.663$ to $0.265~\MeV$, with the common coupling removed.
Its fitted mass coordinate changes from $0.285$ to $0.306$.
Because $\Knf_+^c$ is extracted near $x=0$ in each set, the factor-of-$2.5$ change measures valence-model dependence of the symmetric-point response, separately from the change in the physical mass coordinate.

\section{A possible lattice matching procedure}
\label{sec:lattice}

A finite-volume correlation matrix offers a way to apply the same mode-resolved analysis on the lattice.
We construct a retained conventional block and an orthogonal hybrid block from the interpolators
\begin{equation}
 \{O_{\bar{\bm3}},O_{\bm6},O_{h,+,n},O_{h,-,n}\}.
 \label{eq:lattice_operator_basis}
\end{equation}
Here $O_{\bar{\bm3}}$ and $O_{\bm6}$ create the two conventional flavor structures, and $O_{h,\eta,n}$ denote hybrid interpolators of exchange parity $\eta$.
The mass dependence can be followed along a trajectory near equal light-quark masses~\cite{HorsleyEtAl2011}, using a generalized eigenvalue analysis to extract finite-volume levels and matrix elements~\cite{BlossierEtAl2009,Lang2008}.

\subsection{Orthogonalizing the two operator blocks}

We orthogonalize the hybrid interpolators against the conventional span before normalizing the two blocks.
This ordering preserves the conventional subspace.
Let the positive correlation matrix at reference Euclidean time $t_0$ be
\begin{equation}
 C_0=
 \begin{pmatrix}
 C_{vv}&C_{vh}\\
 C_{hv}&C_{hh}
 \end{pmatrix}.
 \label{eq:lattice_block_metric}
\end{equation}
The subscripts $v$ and $h$ label conventional and hybrid operators, respectively.
All inverses below are taken on a well-conditioned retained subspace.
We first normalize the conventional block:
\begin{equation}
 \widetilde O_v=C_{vv}^{-1/2}O_v.
 \label{eq:lattice_valence_whitening}
\end{equation}
This operation makes its correlation metric the identity; it is also called whitening.
We then subtract the conventional projection from each hybrid operator and normalize the remainder:
\begin{align}
 O_h^\perp&=O_h-C_{hv}C_{vv}^{-1}O_v,
 \nonumber\\
 S_h&=C_{hh}-C_{hv}C_{vv}^{-1}C_{vh},
 \qquad
 \widetilde O_h=S_h^{-1/2}O_h^\perp .
 \label{eq:lattice_hybrid_schur}
\end{align}
The Schur complement $S_h$ is the metric of the remaining hybrid block.
The ordered basis $(\widetilde O_v,\widetilde O_h)$ has unit metric at $t_0$, and its two blocks are orthogonal.
Separate nonsingular changes of the raw conventional and hybrid bases lead only to independent unitary transformations after this procedure.
The singular values of the normalized transition block are therefore unchanged by such basis changes.

Let $\mathsf L$ be the block-triangular transformation defined by Eqs.~\eqref{eq:lattice_valence_whitening} and~\eqref{eq:lattice_hybrid_schur}.
We define
\begin{equation}
 \begin{aligned}
 \widetilde C(t)&=\mathsf L C(t)\mathsf L^\dagger,\\
 H_{\rm eff}(t_0)&=-\frac{1}{a}\log\widetilde C(t_0+a),
 \end{aligned}
 \label{eq:lattice_effective_hamiltonian}
\end{equation}
where $a$ is the Euclidean time spacing and $\widetilde C(t_0)=I$.
The logarithm defines a finite-time effective Hamiltonian in this normalized operator space.
Its conventional--hybrid block supplies a transition matrix $W_{\rm eff}$, and its hybrid diagonal block supplies the intermediate energies.
The reference energy must lie below the included hybrid levels for the positive Gram test.
At least two independent hybrid operators are needed in each exchange sector: with only one row, a transition matrix has no second singular value.

\subsection{Mode and energy truncations}

\begin{figure}[t]
\includegraphics[width=\columnwidth]{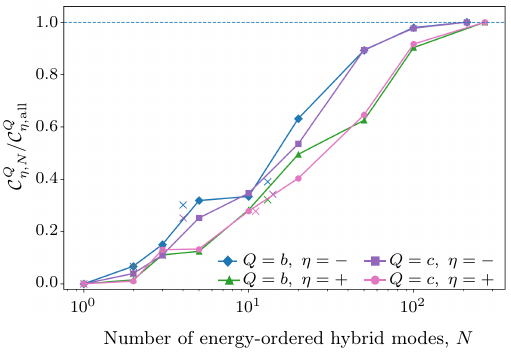}
\caption{Dependence of the nonfactorization coefficient on the included hybrid modes.
Curves show $\Cnf_{\eta,N}^Q/\Cnf_{\eta,\mathrm{all}}^Q$, and crosses in the corresponding sector colors mark excitation cuts of $0.5$, $1$, and $2~\mathrm{GeV}$.
The coefficient accumulates over a broad set of modes in this Gaussian basis.}
\label{fig:mode_saturation}
\end{figure}

A finite lattice operator basis may resolve only a few levels, whereas the constituent model includes many Gaussian pseudostates.
To compare truncated calculations, retain the lowest $N$ intermediate modes in each exchange sector and define
\begin{align}
 G_{\eta,N}&=W_{\eta,N}^\dagger \mathbf D_{\eta,N}^{-1}W_{\eta,N},
 \nonumber\\
 \Cnf_{\eta,N}^Q
 &=\lim_{x\to0}\frac{\gamma_{\eta,N,-}/\gamma_{\eta,N,+}}{x^2}.
 \label{eq:lattice_truncated_C}
\end{align}
The eigenvalues $\gamma_{\eta,N,\pm}$ belong to this truncated Gram matrix.
The different level densities in the two methods motivate using a common excitation-energy cut for comparison.
A second definition uses the excitation energy above the lowest included hybrid state:
\begin{align}
 G_\eta(E_{\max})
 &=\sum_{E_{h_n}-E_{h_1}<E_{\max}}
 \frac{w_{\eta n}^\dagger w_{\eta n}}{D_{\eta n}},
 \nonumber\\
 \Cnf_\eta^Q(E_{\max})
 &=\lim_{x\to0}\frac{\gamma_{\eta,-}(E_{\max})/\gamma_{\eta,+}(E_{\max})}{x^2}.
 \label{eq:lattice_energy_cut_C}
\end{align}
The cut $E_{\max}$ limits the excitation energy $E_{h_n}-E_{h_1}$.
Figure~\ref{fig:mode_saturation} shows the mode and energy-cut dependence; the corresponding values are tabulated in the Supplemental Material.

The first three modes give only $11$--$15\%$ of the full coefficient.
Across the four sectors, the excitation energy of the highest included pseudostate above the lowest mode is \mbox{$4.4$--$4.9~\mathrm{GeV}$} for $N=50$ and \mbox{$9$--$16~\mathrm{GeV}$} for $N=100$.
These high pseudostates serve as Gaussian basis components for the regulated short-distance sum, rather than as individual hybrid-resonance predictions.
The energy-cut calculation gives the same qualitative result: a $2~\mathrm{GeV}$ window includes $11$--$14$ modes but only $28$--$39\%$ of the full coefficient.
The contribution is spread over a broad energy range and depends on the short-distance transition profile, consistent with the effective connected-operator description of Ref.~\cite{BaekEtAl2026PPNP}.

\subsection{Size of the signal}

A second transition direction is detected by a nonzero $\Cnf_{\eta,N}^Q$ or $\Cnf_\eta^Q(E_{\max})$.
A common-form-factor transition gives zero even if its total coupling and mass shift are nonzero.
A nonzero quadratic coefficient demonstrates a second transition direction at that order.
If it vanishes, a second direction may first appear at order $x^6$ or higher in the eigenvalue ratio.

For fixed $Q,\eta$ and $N$, let $s_1\geq s_2\geq0$ be the singular values of $\mathbf D_{\eta,N}^{-1/2}W_{\eta,N}$.
Then $s_2/s_1=\sqrt{\gamma_{\eta,N,-}/\gamma_{\eta,N,+}}$, whose leading small-$x$ estimate is $|x|\sqrt{\Cnf_{\eta,N}^Q}$.
For $N=2$ and $x=0.10$, this estimate is $0.19$--$0.74\%$ across the four sectors.
Using the same leading estimate at the nominal constituent point gives $0.55$--$2.11\%$.
For $N=3$, the corresponding ranges are $0.32$--$1.98\%$ and $0.91$--$5.64\%$.
A calculation would need sub-percent-to-few-percent control of the normalized transition block and several mass points near the symmetric limit.
These are precision targets from the leading small-$x$ term.
An integrated off-diagonal matrix element or an operator-based mixing angle offers an alternative to resolving the smaller singular value.

The dimensional ratios are $3.70$ and $2.62$ in the set A \mbox{spin-$3/2$} model.
In the common-basis spin-complete calculation they are approximately $7.1$ and $7.0$.
The contrast between these ratios makes the heavy-flavor comparison sensitive to the spin content of the intermediate states.

\section{Scope and limitations}
\label{sec:limitations}

The single-direction criterion concerns the full self-energy at a common energy below the intermediate spectrum, for a fixed orthogonal division into retained and hybrid spaces.
Its rank is invariant under basis rotations within those spaces.
The self-energy and hybrid probability also depend on the division between sectors.
For distinct valence energies, the symmetrized second-order operator is Hermitian but need not be positive.
Above an open threshold, the self-energy is complex and the pole problem replaces the positive Gram analysis.

The numerical calculation includes orbital $S$ waves, one singlet color recoupling, and independently propagated $j_{Qg}=1/2,3/2$ sectors.
The minimal vector vertex fixes $\zeta_{1/2}=1$; varying it changes the transition structure, while varying the gaps changes the intermediate dynamics.
The other singlet color component, spin-channel mixing, orbital excitations, and noncentral hybrid forces are omitted.
The reference three-quark shifts fix an overall strength rather than a microscopic operator matching: the full self-energy includes one- and two-body pieces that have been subtracted from the connected reference interaction.
The $6\%$ basis estimate and $9$--$17\%$ Gaussian-to-Yukawa variation quantify numerical and radial-profile sensitivity, not the full uncertainty from these omitted structures.
The basis estimate applies to integrated coefficients, not individual high pseudostate energies.

A model-to-lattice comparison must match the operator subspaces, heavy-field normalization, and light-quark mass trajectory.
Our source and mass-operator constructions are nonrelativistic analogues of the lattice definitions.
The finite-time Hamiltonian in Eq.~\eqref{eq:lattice_effective_hamiltonian} also depends on $t_0$, lattice spacing, and volume; equality with the exact projected QCD Hamiltonian is not assumed.
The small-$x$ precision targets omit higher-order mass dependence and lattice statistical and systematic errors.
The coefficients in Eq.~\eqref{eq:matched_local_operator} parametrize a two-state matrix using the pure-spin matrices of Appendix~\ref{app:spin}; projecting microscopic spin operators onto spatial states requires their overlap factors and basis transformations.

Radiative and semileptonic amplitudes remain to be evaluated with the effective currents derived here.
The hybrid probability supplies a normalization scale, but a suppressed leading amplitude or a transition current can produce a larger relative correction.
It therefore does not by itself bound a decay rate or a decay-extracted angle.

\section{Conclusion}

The probability, source-overlap, and mass-operator definitions give different mixing angles for the same constituent-quark wave functions.
For charm, the probability angle is $3.11^\circ$ in set A, while the operator-based angles are about $2.5^\circ$ in set A and $2.4^\circ$ in set B.
Changing the definition thus brings the model values closer to the lattice determinations.

The minimal spin-complete chromomagnetic hybrid contribution acts in the opposite direction.
After normalization to the reference short-range mass shifts, it raises the source and mass-operator charm angles to $2.67$--$3.16^\circ$, with a hybrid probability below $1.7\%$.
The bottom operator-based angles remain of order $0.3$--$0.5^\circ$.
In this color-triplet orbital-$S$-wave realization, virtual hybrids therefore increase the charm-angle difference from the lattice results.

The mode-resolved analysis explains why a change in the mixing angle and the emergence of a second transition direction are separate effects.
The parallel flavor-breaking response controls the off-diagonal mass term, while the perpendicular response measures nonfactorization.
A rank-one correction occurs only when all weighted transitions select the same retained-state combination.
The energy-cut sums show that the second direction develops over a broad range of intermediate modes.
Together with the spin-channel and heavy-flavor comparisons, this identifies the transition structures that are lost in a single-form-factor approximation.
The effective-current relation provides the next step from these mass and wave-function results to decay amplitudes.
\raggedbottom
\begin{acknowledgments}
S.K. acknowledges support from the ANCHOR program through the Gyeongbuk ANCHOR CENTER, funded by the Ministry of Education (MOE) and Gyeongsangbuk-do, Republic of Korea (No.~\mbox{2026-ANCHOR-15-104}).
J.B. acknowledges support from the National Research Foundation of Korea (NRF) under Grant No.~\mbox{RS-2023-NR077232}.

\end{acknowledgments}

\appendix

\Needspace{12\baselineskip}
\section{Spin recouplings}
\label{app:spin}

In the ordered pure-spin basis $(|\rho\rangle,|\lambda\rangle)$ of Eq.~\eqref{eq:valence_spin_basis}, the pair-spin matrices are
\begin{align}
 \bm\sigma_q\cdot\bm\sigma_s
 &=\begin{pmatrix}-3&0\\0&1\end{pmatrix},
 \nonumber\\
 \bm\sigma_q\cdot\bm\sigma_Q
 &=\begin{pmatrix}0&-\sqrt3\\-\sqrt3&-2\end{pmatrix},
 \nonumber\\
 \bm\sigma_s\cdot\bm\sigma_Q
 &=\begin{pmatrix}0&+\sqrt3\\+\sqrt3&-2\end{pmatrix}.
 \label{eq:spin_matrices_appendix}
\end{align}
These matrices act between pure spin functions; projection onto spatial wave functions supplies the corresponding overlap factors.
They also define the effective two-state spin-matrix convention used for the coefficients in Eq.~\eqref{eq:matched_local_operator}.
The \mbox{spin-$3/2$} transition gives Eq.~\eqref{eq:j32_spin_rows}.
For spin $1/2$, the Clebsch--Gordan construction gives $\mathsf R_\ell^\dagger=\sigma_{Q,\ell}/\sqrt3$ and Eq.~\eqref{eq:jhalf_recoupling}.
An independent tensor calculation reproduces the transition color factor $-2/\sqrt3$ and the six pair color factors in Eq.~\eqref{eq:hybrid_color_factors}.

\section{Numerical checks}
\label{app:technical}

The direct determinant and the Cauchy--Binet sum agree to $1.5\times10^{-14}$ or better over the sampled mass points.
We also solve a finite detuning problem, described in the Supplemental Material, whose small positive detuning root approaches $(\gtr)^2\gamma_-$.
At unit reduced strength, their largest relative difference is $1.4\times10^{-5}$.
This comparison checks the implementation of the reduction.
The Supplemental Material gives the basis, regulator, central-interaction, spin-channel, sector-offset, and nondegenerate-denominator comparisons.

\section*{Data availability}
The numerical results reported in this work are presented in the article and its Supplemental Material.

\end{document}


\title{Supplemental Material for ``Mass and decay angles in the \texorpdfstring{$\Xi_c$--$\Xi_c'$}{Xi-c--Xi-c-prime} mixing tension: Virtual hybrids and a single-direction criterion''}
\author{Sanghoon Kim}
\affiliation{Department of Computer Education, Gyeongkuk National University, Andong 36729, South Korea}

\author{Gihwan Nam}
\affiliation{Department of Physics, Yonsei University, Seoul 03722, South Korea}

\author{Jongheon Baek}
\email[Corresponding author: ]{bjh6941@yonsei.ac.kr}
\affiliation{Department of Physics and Institute of Physics and Applied Physics, Yonsei University, Seoul 03722, Korea}
\date{October 7, 2026}
\maketitle
\raggedbottom

\section{Conventions and the determinant identity}
\enlargethispage{10pt}

We use the same conventions as in the main text.
The heavy-quark label is $Q=c,b$; the sans-serif projectors $\mathsf P$ and $\mathsf Q$ select the two retained valence states and the intermediate hybrid space, respectively.
The reduced transition operator $T:\mathsf P\to\mathsf Q$ and its matrix $W_{n\alpha}=\langle h_n|T|v_\alpha\rangle$ are defined with the common dimensionless strength $\gtr$ factored out.
The physical off-diagonal Hamiltonian is $\gtr T$, and its eliminated-sector contribution is $\Sigma=-(\gtr)^2G$.
For a reference energy $E$ below the intermediate spectrum, $D_n=E_{h_n}-E>0$ and $G=W^\dagger \mathbf D^{-1}W$.
Thus $W$, $G$, their eigenvalues $\gamma_\pm$, and the dimensional coefficient $\Knf_\eta^Q$ have units of MeV, whereas $\Cnf_\eta^Q$ is dimensionless.
Numerical entries use this reduced convention unless a value of $\gtr$ is given.

Let $A=\mathbf D^{-1/2}W$ be the weighted matrix with two columns, and denote its rows by $a_n^T$.
The Cauchy--Binet identity is
\begin{equation}
 \det(A^\dagger A)=\sum_{m<n}|\det(a_m,a_n)|^2.
 \label{eq:s_cauchy_binet}
\end{equation}
Each term compares two transition rows.
If $G\neq0$, the sum vanishes precisely when all nonzero rows are collinear, giving rank one.
The operator expression $G_{\alpha\beta}=\langle v_\alpha|T^\dagger(H_{\mathsf Q\mathsf Q}-E)^{-1}T|v_\beta\rangle$ makes its invariance under basis changes within a fixed intermediate space explicit.
Equation~\eqref{eq:s_cauchy_binet} is its representation in intermediate energy eigenstates.

Near the symmetric point, write the exchange-even columns as
\begin{equation}
 A(x)=\bigl[xb+x^3b_3+\cdots,\ a+x^2a_2+\cdots\bigr],
 \label{eq:s_column_expansion}
\end{equation}
where $a\neq0$ is the allowed column at $x=0$, and $b$ is the first response of the forbidden column.
The expansion coordinate is $x=(m_s-m_q)/(m_s+m_q)$ at fixed $(m_q+m_s)/2$.
The column norms and inner product give
\begin{align}
 G_{11}&=x^2\langle b|b\rangle+O(x^4),
 &G_{22}&=\langle a|a\rangle+O(x^2),
 \nonumber\\
 G_{12}&=x\langle b|a\rangle+O(x^3),
 \label{eq:s_gram_expansion}
\end{align}
and
\begin{equation}
 \begin{aligned}
 \gamma_-(x)&=x^2\left(\langle b|b\rangle-
 \frac{|\langle a|b\rangle|^2}{\langle a|a\rangle}\right)+O(x^4)\\
 &=x^2\|P_a^\perp b\|^2+O(x^4).
 \end{aligned}
 \label{eq:s_small_eigenvalue}
\end{equation}
Here $P_a^\perp=I-|a\rangle\langle a|/\langle a|a\rangle$ removes the parallel component.
The exchange-odd result follows by interchanging the columns.
An ordinary first-order valence rotation adds $\theta'_{\rm conv}a$ to $b$, but
\begin{equation}
 P_a^\perp(b+\theta'_{\rm conv}a)=P_a^\perp b,
 \label{eq:s_rotation_cancellation}
\end{equation}
so the leading nonfactorization coefficient is unchanged.
It measures the part of the transition response that cannot be absorbed into that common rotation.

We also test a Hamiltonian limit in which factorization is exact.
At $x=0$, we solve the spin-independent central valence Hamiltonian and use its ground spatial orbital $\phi_0$ for both spin states.
We keep this orbital, the hybrid states, their denominators, and the radial transition shape fixed, while varying only
\begin{equation}
 f_i(x)=\frac{\mu_{iQ}(x)\mu'_{iQ}(x)}{m_i(x)m_Q}.
\end{equation}
The two light-quark spatial overlaps then differ by the exchange sign $\eta$.
For the even and odd sectors, respectively,
\begin{align}
 W_{+n}&\propto R_n
 \left(\sqrt2[f_q-f_s],-\sqrt{\frac23}[f_q+f_s]\right),
 \nonumber\\
 W_{-n}&\propto R_n
 \left(\sqrt2[f_q+f_s],-\sqrt{\frac23}[f_q-f_s]\right),
\end{align}
where $R_n$ is a mode-dependent radial overlap.
Within each sector, the rows select the same retained-state combination.
All minors vanish and $\gamma_-=0$.
The numerical coefficients are below $2\times10^{-15}$ in all four heavy-flavor and exchange-parity sectors.

\subsection{Valence spectrum cross-check}
\label{sec:s_valence_spectrum}

The valence masses reported in the main text are obtained from its displayed Hamiltonian and parameter table, with the common transition strength set to zero.
For each of the three Jacobi rearrangements, we use physical relative coordinates
\[
 \bm r_{ij}=\bm r_i-\bm r_j,\qquad
 \bm R_{k,(ij)}=\bm r_k-\frac{m_i\bm r_i+m_j\bm r_j}{m_i+m_j},
\]
and Gaussian factors $\exp[-r_{ij}^2/r_n^2-R_{k,(ij)}^2/R_N^2]$.
The seven ranges in each coordinate form a geometric sequence from $0.05$ to $3.20~\mathrm{fm}$.
Both pair-spin functions are included, giving $3\times7^2\times2=294$ basis vectors before overlap orthogonalization.
All color-singlet pair factors are $\langle F_i\cdot F_j\rangle=-2/3$, and the kinetic operator has the center-of-mass motion removed.
This independent evaluation gives $(2489.6430,2585.3954)~\MeV$ for charm and $(5850.1757,5977.8700)~\MeV$ for bottom.
The corresponding lower-state norm angles are $3.113584^\circ$ and $0.894674^\circ$, which reproduce the $3.1136^\circ$ and $0.8947^\circ$ results and the $95.75$ and $127.69~\MeV$ splittings in this basis.

\section{Comparison of angle definitions}

We compare a spin-probability angle with two operator-based angles, using the same model wave functions.
The nonnegative norm angle is $\theta_N=\tan^{-1}\sqrt{P_{\rm sub}/P_{\rm dom}}$, where $P_{\rm sub}$ and $P_{\rm dom}$ are the smaller and larger spin-component probabilities.
The source-overlap angle $\theta_{\rm src}(R)$ is obtained by fitting the two-state overlap matrix with independent row normalizations and a common rotation.
We vary the common Gaussian source radius over $0.1\leq R\leq1.0~\mathrm{fm}$.

For the mass-operator angle, define $m_0=(2m_q^{\rm fit}+m_s^{\rm fit})/3$ from the fitted masses and first solve for the two flavor eigenstates at $m_q=m_s=m_0$.
The physical Hamiltonian is then projected between these reference states and diagonalized to obtain $\theta_H$.
This is distinct from the rank-test trajectory, whose symmetric mass is $\bar m=(m_q^{\rm fit}+m_s^{\rm fit})/2$.
Table~\ref{tab:s_angle_definitions} gives the detailed values; the corresponding comparison is shown in Fig.~2 of the main text.

\begin{table*}[h]
\footnotesize
\caption{Values of the three model angles.
The spin-complete ranges use $670\leq\Delta_{1/2}\leq1070~\MeV$ and $-400\leq\delta_-\leq600~\MeV$ with the one-strength reference normalization.
Leakage is the probability outside the two SU(3)-symmetric reference states.
Sets A and B use the two valence parameter sets defined in Sec.~I of the main text.}
\label{tab:s_angle_definitions}
\label{tab:angle_definitions}
\begin{ruledtabular}
\setlength{\tabcolsep}{4pt}
\begin{tabular}{@{}l c c c c@{}}
case&$\theta_N$&$\theta_{\rm src}(0.4~\mathrm{fm})$&$\theta_H$&\shortstack{two-state\\leakage}\\
\hline
set A $\Xi_c$&$3.1136^\circ$&$2.5040^\circ$&$2.4846^\circ$&$1.37\%$\\
set B $\Xi_c$&$2.8083^\circ$&$2.4110^\circ$&$2.3809^\circ$&$1.63\%$\\
set A minimal spin complete, charm&$3.268$--$3.662^\circ$&$2.693$--$3.158^\circ$&$2.671$--$3.135^\circ$&--\\
set A $\Xi_b$&$0.8947^\circ$&$0.4490^\circ$&$0.4753^\circ$&$1.74\%$\\
set B $\Xi_b$&$0.6868^\circ$&$0.3363^\circ$&$0.3595^\circ$&$2.06\%$\\
set A minimal spin complete, bottom&$0.892$--$0.908^\circ$&$0.423$--$0.454^\circ$&$0.450$--$0.480^\circ$&--
\end{tabular}
\end{ruledtabular}
\end{table*}

The source-radius intervals are $2.4807$--$2.5812^\circ$ for charm with set A, $2.3848$--$2.4897^\circ$ for charm with set B, $0.4320$--$0.5328^\circ$ for bottom with set A, and $0.3155$--$0.4204^\circ$ for bottom with set B.
The residual of the fitted overlap rotation is below $2.7\times10^{-5}$.
For charm with set A, changing from the norm angle to the operator-based definitions reduces the angle by about $0.5$--$0.7^\circ$.
The remaining model values are still above the lattice determinations quoted in the main text.

\section{Sources of flavor-breaking nonfactorization}

For each heavy flavor and exchange parity, we retain the well-conditioned symmetry subspace selected at $x=0$ and fit
\begin{equation}
 y(x)=a_2x^2+a_4x^4,
 \qquad 0.001\leq x\leq0.008,
 \label{eq:s_fit_form}
\end{equation}
with zero intercept.
The choice $y=\gamma_-$ gives $\Knf_\eta^Q=a_2$, and we use $\Cnf_\eta^Q=\Knf_\eta^Q/\gamma_{\eta,+}(0)$.
Table~\ref{tab:s_mechanisms} compares five controlled approximations with the full calculation.
The common-orbital approximation uses the same symmetric-point spatial state for both spin functions and varies the scalar mass factors $f_i(x)$.
The mass-factor approximation keeps distinct symmetric-point orbitals and varies only $f_i(x)$.
For the vertex response, we vary the full radial vertex with the states fixed at $x=0$.
For the state response, we vary the valence states, hybrid states, and excitation denominators with the vertex fixed at $x=0$.
The fixed-range approximation includes the state response and scalar mass factors while holding the Gaussian transition range fixed.
The coefficient is quadratic in the perpendicular response vector, so combining the responses introduces interference terms.

\begin{table*}[h]
\caption{Controlled approximations to the set A \mbox{spin-$3/2$} result.
The five controls are defined in the text.
The coefficient $\Knf_\eta^Q$ is in MeV before multiplication by $(\gtr)^2$.}
\label{tab:s_mechanisms}
\label{tab:mechanism_decomposition}
\scriptsize
\resizebox{\textwidth}{!}{%
\begin{ruledtabular}
\begin{tabular}{c c c c c c c c c}
$Q$&$\eta$&\shortstack{common\\orbital}&\shortstack{mass\\factors}&\shortstack{vertex\\response}&\shortstack{state\\response}&\shortstack{fixed\\range}&full $\Cnf_\eta^Q$&full $\Knf_\eta^Q$\\
\hline
$c$&$+$&$<2\times10^{-15}$&$0.002411$&$0.08014$&$0.10136$&$0.12640$&$0.22374$&$0.66285$\\
$c$&$-$&$<2\times10^{-15}$&$0.0002493$&$0.009379$&$0.016891$&$0.013405$&$0.009324$&$0.035879$\\
$b$&$+$&$<2\times10^{-15}$&$0.002205$&$0.12480$&$0.14554$&$0.17774$&$0.35310$&$0.17919$\\
$b$&$-$&$<2\times10^{-15}$&$0.0002013$&$0.014791$&$0.030959$&$0.026519$&$0.018543$&$0.013695$
\end{tabular}
\end{ruledtabular}}
\end{table*}

Varying the mass factors with distinct $\rho$ and $\lambda$ orbitals gives $0.6$--$2.7\%$ of the full coefficient.
The vertex response gives $35$--$36\%$ of the even coefficients and $80$--$101\%$ of the odd coefficients.
Allowing only the state and denominator response gives $41$--$45\%$ of the even result and $167$--$181\%$ of the odd result.
Values above the full odd-sector result indicate destructive interference when the effects are combined.
With the Gaussian width fixed, the odd coefficients are $43$--$44\%$ larger than the full values.
The range dependence partly cancels the other contributions in the odd sectors, while the even sectors show constructive interference.

\section{Independent color and spin checks}

The retained color-singlet component is
\begin{equation}
 |[(qs)_{\bar{\bm3}_C}(Qg)_{\bm3_C}]_{\bm1_C}\rangle.
\end{equation}
An independent tensor calculation gives the pair color factors
\begin{equation}
\begin{array}[b]{c|rrrrrr}
 ij & qs&qQ&sQ&qg&sg&Qg\\ \hline
 \langle F_i\cdot F_j\rangle
 &-2/3&1/12&1/12&-3/4&-3/4&-3/2
\end{array}.
\end{equation}
Their sum is $-7/2$, and the projected transition color factor is $-2/\sqrt3$.
The propagated space is restricted to this singlet; coupling to the orthogonal $[(qs)_{\bm6_C}(Qg)_{\overline{\bm6}_C}]_{\bm1_C}$ component is discussed below in relation to the emission map.

For $j_{Qg}=3/2$, direct multiplication of the transition matrices gives
\begin{equation}
 \langle[\tfrac12_i\otimes\tfrac32_{Q^*}]_{s'}|
 \bm\sigma_i\cdot\bm{\mathsf S}_Q^\dagger|
 [\tfrac12_i\otimes\tfrac12_Q]_{s}\rangle
 =\frac{2\sqrt6}{3}\delta_{s1}\delta_{s'1}.
\end{equation}
Here $s$ and $s'$ are the initial and final pair spins; equal magnetic projections are implicit.
The symbol $Q^*$ labels the coupled $Qg$ spin state.
Recoupling to the initial $(\rho,\lambda)$ basis gives
\begin{equation}
 qQ:\ (\sqrt2,-\sqrt{2/3}),\qquad
 sQ:\ (-\sqrt2,-\sqrt{2/3}).
\end{equation}
For $j_{Qg}=1/2$, the Clebsch--Gordan construction gives
\begin{equation}
 \mathsf R_x^\dagger=\frac{\sigma_x}{\sqrt3},\qquad
 \mathsf R_y^\dagger=\frac{\sigma_y}{\sqrt3},\qquad
 \mathsf R_z^\dagger=\frac{\sigma_z}{\sqrt3},
 \label{eq:s_jhalf_matrices}
\end{equation}
and the local scalar transition is $(1/\sqrt3)\bm\sigma_i\cdot\bm\sigma_Q$.
With final light-pair spins $s_{qs}=0,1$ as rows and initial $\rho,\lambda$ as columns,
\begin{equation}
 M_q^{(1/2)}=
 \begin{pmatrix}0&-1\\-1&-2/\sqrt3\end{pmatrix},
 \qquad
 M_s^{(1/2)}=
 \begin{pmatrix}0&+1\\+1&-2/\sqrt3\end{pmatrix}.
 \label{eq:s_jhalf_recoupling}
\end{equation}
The radial function $u_{iQ}$ has units of $\MeV^3$.
Including the inverse-mass factors gives a reduced vertex with units of MeV.
Thus $W$ and $G$ have energy units, and $\gamma_-/\gamma_+$ is dimensionless.
The physical self-energy is $-(\gtr)^2G$, as in Sec.~I.

\section{Small-\texorpdfstring{$x$}{x} coefficients}

\begin{figure}[h]
\includegraphics[width=0.92\textwidth]{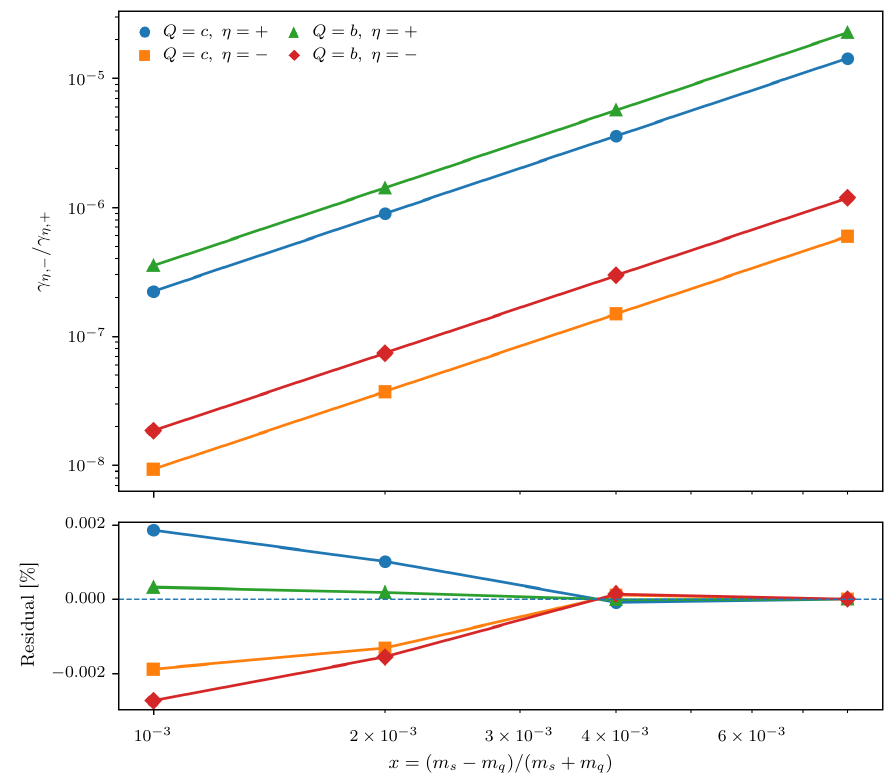}
\caption{Fits of $\gamma_{\eta,-}/\gamma_{\eta,+}$ to $c_2x^2+c_4x^4$ in the four set A \mbox{spin-$3/2$} sectors.
The lower panels show the relative residuals.
The allowed leading power follows from exchange symmetry; the nonzero coefficient is the dynamical result.}
\label{fig:s_small_x_fit}
\end{figure}

\begin{figure}[h]
\includegraphics[width=0.75\textwidth]{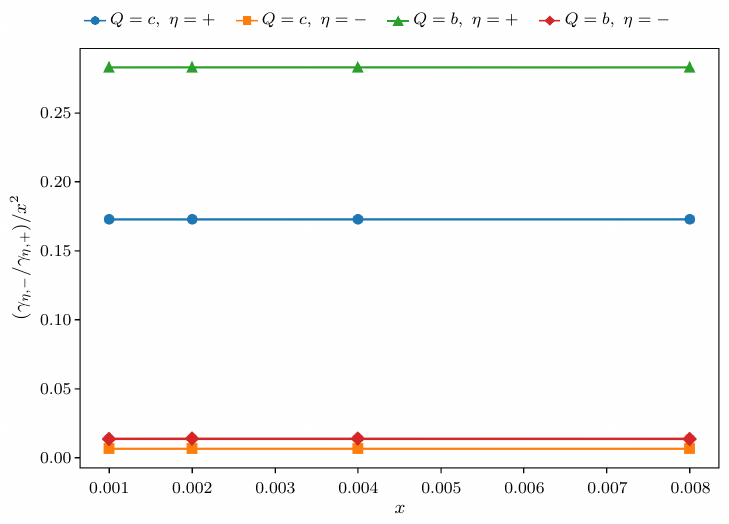}
\caption{Small-$x$ behavior of $(\gamma_{\eta,-}/\gamma_{\eta,+})/x^2$ in the common 10-triple basis used for Table~\ref{tab:s_jhalf}.
These curves illustrate the limiting behavior.
Table~\ref{tab:s_coefficients} gives the corresponding coefficients from 15 range triples per tree.}
\label{fig:s_plateau}
\end{figure}

The 15-triple basis contains 540 correlated Gaussians before overlap reduction.
At $x=0$, orthogonalization retains 270 even and 210 odd directions.
We keep these dimensions fixed along the light-flavor trajectory rather than applying a new rank selection at every mass point.
This procedure avoids apparent changes caused by overlap eigenvalues crossing the numerical cutoff.
The coefficients in Table~\ref{tab:s_coefficients} use Eq.~\eqref{eq:s_fit_form}.
A direct fit to $\gamma_-/\gamma_+$ gives the same $\Cnf_\eta^Q$ to within a relative difference of $9.3\times10^{-5}$ in all sectors.
The reported coefficients obey $\Cnf_\eta^Q=\Knf_\eta^Q/\gamma_{\eta,+}(0)$.
Figure~\ref{fig:s_small_x_fit} shows the fits and their relative residuals.
Figure~\ref{fig:s_plateau} shows the limiting ratios in the common 10-triple basis used for the spin-channel comparison.

\begin{table}[h]
\caption{Set A coefficients from 15 range triples per Jacobi tree.
The eigenvalue $\gamma_{\eta,+}(0)$ and coefficient $\Knf_\eta^Q$ are in MeV, with the common strength removed.
The coefficient $\Cnf_\eta^Q$ is dimensionless.}
\label{tab:s_coefficients}
\begin{ruledtabular}
\begin{tabular}{c c c c c}
$Q$&$\eta$&$\gamma_{\eta,+}(0)~[\MeV]$&$\Cnf_\eta^Q$&$\Knf_\eta^Q~[\MeV]$\\
\hline
$c$&$+$&$2.9626$&$0.22374$&$0.66285$\\
$c$&$-$&$3.8480$&$0.009324$&$0.035879$\\
$b$&$+$&$0.50748$&$0.35310$&$0.17919$\\
$b$&$-$&$0.73854$&$0.018543$&$0.013695$
\end{tabular}
\end{ruledtabular}
\end{table}

The dimensional charm-to-bottom ratios are
\begin{equation}
 \Knf_+^c/\Knf_+^b=3.699,
 \qquad
 \Knf_-^c/\Knf_-^b=2.620,
\end{equation}
and the even-to-odd ratios are $18.48$ for charm and $13.08$ for bottom.
We use the same reduced strength $\gtr$ for charm and bottom.

\section{Charm-to-bottom ratios}

At the symmetric point, the explicit heavy-spin factor and the radial normalization give
\begin{equation}
 \left(\frac{m_b}{m_c}\right)^2=7.9936,
 \qquad
 \left(\frac{\mu_c\mu_c'}{\mu_b\mu_b'}\right)^2=0.81625.
\end{equation}
Here $\mu_Q$ and $\mu_Q'$ abbreviate the light--heavy regulator scales at the same symmetric light mass.
We repeat the calculation with every intermediate denominator set to $870~\MeV$ to separate the overlap and denominator effects.
Table~\ref{tab:s_hq_decomp} gives the resulting multiplicative comparison.

\begin{table}[h]
\caption{Multiplicative decomposition of $\Knf_\eta^c/\Knf_\eta^b$ in the $j_{Qg}=3/2$ model.}
\label{tab:s_hq_decomp}
\begin{ruledtabular}
\begin{tabular}{l c c}
factor&$\eta=+$&$\eta=-$\\
\hline
$1/m_Q^2$&$7.9936$&$7.9936$\\
$\mu\mu'$ correction&$0.81625$&$0.81625$\\
spatial overlap&$0.5972$&$0.4292$\\
denominator spectrum&$0.9493$&$0.9356$\\
\hline
product&$3.699$&$2.620$
\end{tabular}
\end{ruledtabular}
\end{table}

The common-denominator charm-to-bottom ratios are $3.897$ and $2.800$.
Restoring the intermediate spectrum changes them moderately.
Spatial overlaps give the largest correction beyond the explicit mass and radial-normalization factors.

\section{Spin and color content of the minimal vertex}

A spin-one gluonic mode and a spin-$1/2$ quark have complementary spin projectors:
\begin{align}
 \Pi^{(3/2)}_{\ell\ell'}&=\delta_{\ell\ell'}I-\frac13\sigma_{Q,\ell}\sigma_{Q,\ell'},&
 \Pi^{(1/2)}_{\ell\ell'}&=\frac13\sigma_{Q,\ell}\sigma_{Q,\ell'}.
\end{align}
Here $\ell,\ell'=x,y,z$, $\sigma_{Q,\ell}$ are the heavy-quark Pauli matrices, and $I$ is the identity in its two-dimensional spin space.
The matrices $\Pi^{(j)}_{\ell\ell'}$ are Cartesian blocks of the total-spin projectors.
They satisfy $\Pi^{(3/2)}_{\ell\ell'}+\Pi^{(1/2)}_{\ell\ell'}=\delta_{\ell\ell'}I$.
The independent tensor implementation reproduces this identity to $2.3\times10^{-16}$.
For the bare vector-creation map, both spin sectors inherit one common reduced coefficient.
We denote their relative transition amplitude by $\zeta_{1/2}$, with $\zeta_{1/2}=1$ for this minimal vertex.
A general effective transition admits independent reduced coefficients under the Wigner--Eckart theorem; varying $\zeta_{1/2}$ tests these additional structures.

The normalized color emission map is
\begin{equation}
 V|\xi\rangle=\frac{\sqrt3}{2}\sum_{a,\eta}(F^a)_{\eta\xi}|q_\eta g_a\rangle,
\end{equation}
where $\xi,\eta$ are quark color indices and $a$ is a gluon color index.
Its image is the unique $\mathbf3_C$ component of $\mathbf3_C\otimes\mathbf8_C$.
The numerical color projectors give
\begin{equation}
 \|P_{\overline{\mathbf 6}}V\|<10^{-15},
 \qquad
 \|P_{\mathbf{15}}V\|<4\times10^{-16}.
\end{equation}
The minimal map therefore excites only the first singlet directly.
The present calculation propagates that component; adding the second requires color mixing in $\mathsf Q$ or another transition operator.

\section{The spin-\texorpdfstring{$1/2$}{1/2} contribution}

The propagated spin-$1/2$ and \mbox{spin-$3/2$} sectors are taken to be orthogonal and dynamically uncoupled.
Their self-energies therefore add, and the relative phase cancels in the sum.
With $\zeta_{1/2}$ multiplying the spin-$1/2$ transition amplitude, the reduced Gram matrix is $G=G_{3/2}+\zeta_{1/2}^2G_{1/2}$.
The minimal bare vector vertex uses $\zeta_{1/2}=1$.
Changing this value tests the transition operator, whereas varying the gaps tests the diagonal hybrid Hamiltonian.
Table~\ref{tab:s_jhalf} uses a common 10-triple-per-tree basis and the fixed-power fit of Eq.~\eqref{eq:s_fit_form}.

\begin{table*}[h]
\caption{Spin-channel comparison in a common basis.
The dimensional coefficient $\Knf_\eta^Q$ is in MeV with $(\gtr)^2$ removed.
The combined column uses the minimal relative amplitude and equal reference gaps.}
\label{tab:s_jhalf}
\label{tab:jhalf}
\begin{ruledtabular}
\begin{tabular}{c c cc cc cc}
&&\multicolumn{2}{c}{$j_{Qg}=3/2$}&\multicolumn{2}{c}{$j_{Qg}=1/2$}&\multicolumn{2}{c}{combined}\\
$Q$&$\eta$&$\Cnf_\eta^Q$&$\Knf_\eta^Q$&$\Cnf_\eta^Q$&$\Knf_\eta^Q$&$\Cnf_\eta^Q$&$\Knf_\eta^Q$\\
\hline
$c$&$+$&$0.1728$&$0.5071$&$0.0759$&$0.7078$&$0.4517$&$5.5379$\\
$c$&$-$&$0.00649$&$0.02402$&$0.7459$&$1.7261$&$0.3010$&$1.8105$\\
$b$&$+$&$0.2829$&$0.1459$&$0.0851$&$0.1293$&$0.3846$&$0.7827$\\
$b$&$-$&$0.01371$&$0.01008$&$0.6137$&$0.2472$&$0.2285$&$0.2601$
\end{tabular}
\end{ruledtabular}
\end{table*}

The large odd-sector response follows from Eq.~\eqref{eq:s_jhalf_recoupling}.
A final light-pair spin-zero state has no local transition from the initial $\rho$ state and receives opposite-sign contributions from the two paths out of $\lambda$.
For total exchange parity $-$, this spin channel has a spatially symmetric orbital.
The $\lambda$ amplitude cancels at $x=0$, but its first flavor-breaking response enters the forbidden column without an accompanying allowed-$\rho$ component.
It therefore contributes efficiently to $P_a^\perp b$.
For total exchange parity $+$, the same spin channel has a spatially antisymmetric orbital and a smaller effect.

In the same basis, the dimensional heavy-flavor ratios are
\begin{equation}
\begin{array}[b]{c|cc}
&\eta=+&\eta=-\\ \hline
j_{Qg}=3/2&3.48&2.38\\
j_{Qg}=1/2&5.48&6.98\\
\text{combined}&7.08&6.96
\end{array}.
\end{equation}
The ratio reflects the retained spin content as well as the explicit heavy-mass prefactor.

For each spin sector $i$, write $b_i=\beta_i a_i+b_i^\perp$, with $\beta_i=\langle a_i|b_i\rangle/\|a_i\|^2$.
Define $n_i=\|a_i\|^2$ and $\Knf_i=\|b_i^\perp\|^2$.
Projection perpendicular to the combined allowed vector gives
\begin{equation}
 \Knf_{3/2\oplus1/2}=\Knf_{3/2}+\Knf_{1/2}
 +\frac{n_{3/2}n_{1/2}}{n_{3/2}+n_{1/2}}
 |\beta_{3/2}-\beta_{1/2}|^2.
 \label{eq:s_spin_mismatch}
\end{equation}
Table~\ref{tab:s_beta_decomposition} verifies this identity numerically.

\begin{table*}[h]
\caption{Parallel and perpendicular contributions in the two spin sectors.
The values $n_i=\gamma_{+,i}(0)$ and all $\Knf_\eta^Q$ columns are in MeV before multiplication by $(\gtr)^2$.
The slopes $\beta_i$ are real and dimensionless.}
\label{tab:s_beta_decomposition}
\label{tab:spin_channel_mismatch}
\scriptsize
\resizebox{\textwidth}{!}{%
\begin{ruledtabular}
\begin{tabular}{c c r r r r r r r r}
$Q$&$\eta$&$n_{3/2}$&$n_{1/2}$&$\beta_{3/2}$&$\beta_{1/2}$&$\Knf_{3/2}$&$\Knf_{1/2}$&$\Knf_{\rm mis}$&$\Knf_{\rm comb}$\\
\hline
$c$&$+$&$2.935$&$9.326$&$+0.866$&$-0.525$&$0.507$&$0.708$&$4.323$&$5.538$\\
$c$&$-$&$3.700$&$2.314$&$+0.181$&$-0.024$&$0.0240$&$1.726$&$0.0604$&$1.810$\\
$b$&$+$&$0.516$&$1.520$&$+0.819$&$-0.329$&$0.146$&$0.129$&$0.5075$&$0.7827$\\
$b$&$-$&$0.735$&$0.403$&$+0.0169$&$-0.0859$&$0.0101$&$0.247$&$0.00275$&$0.2601$
\end{tabular}
\end{ruledtabular}}
\end{table*}

The mismatch of the two parallel slopes gives $78.1\%$ and $64.8\%$ of the even charm and bottom coefficients.
It gives only $3.3\%$ and $1.1\%$ in the odd sectors, where the intrinsic spin-$1/2$ perpendicular response dominates.
The weighted mean slope $\bar\beta=(\sum_i n_i\beta_i)/(\sum_i n_i)$ controls the leading off-diagonal element for the real amplitudes used here.
Equation~\eqref{eq:s_spin_mismatch} measures a different quantity: the component perpendicular to the combined allowed column.

\begin{figure}[h]
\includegraphics[width=0.75\textwidth]{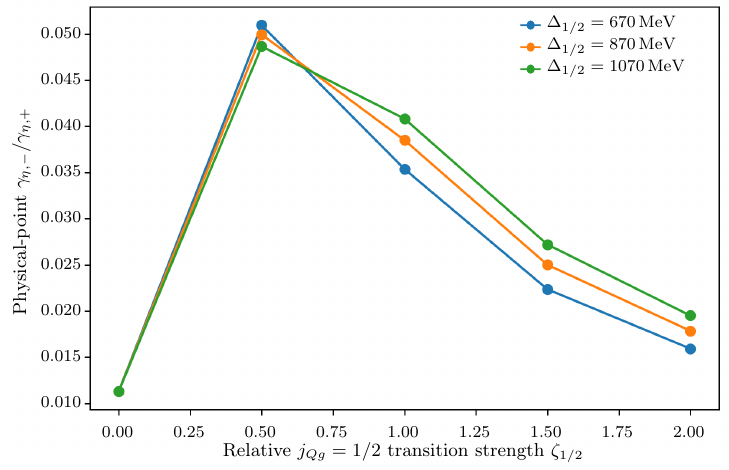}
\caption{Charm-even eigenvalue-ratio dependence on the relative spin-$1/2$ transition amplitude and reference gap at the nominal fitted mass point.}
\label{fig:s_jhalf_scan}
\end{figure}

We also vary $0\leq\zeta_{1/2}\leq2$ and $670\leq\Delta_{1/2}\leq1070~\MeV$ at the nominal fitted mass point.
The smaller-to-larger eigenvalue ratio is $0.0113$--$0.0509$ for charm-even and $4.67\times10^{-4}$--$0.0521$ for charm-odd.
The bottom ranges are $0.0147$--$0.0434$ and $8.20\times10^{-4}$--$0.0402$.
Figure~\ref{fig:s_jhalf_scan} shows the charm-even slice.

\section{Mode and energy truncations}

To study how many intermediate states contribute, retain the lowest $N$ modes and form
\begin{equation}
 G_{\eta,N}=W_{\eta,N}^\dagger \mathbf D_{\eta,N}^{-1}W_{\eta,N},
 \qquad
 \Cnf_{\eta,N}^Q=\lim_{x\to0}
 \frac{\gamma_{\eta,N,-}/\gamma_{\eta,N,+}}{x^2}.
\end{equation}
The eigenvalues $\gamma_{\eta,N,\pm}$ refer to this truncated matrix.
Modes are ordered by their energies at each $x$ within the transported exchange subspace.
Table~\ref{tab:s_mode_truncation} gives the coefficients and the excitation energy of the highest included mode relative to the lowest mode.

\begin{table*}[h]
\caption{Mode-truncated coefficients and excitation energies. Each coefficient column gives $\Cnf_{\eta,N}^Q$; $\Delta E_N=E_N-E_1$ is in GeV.}
\label{tab:s_mode_truncation}
\label{tab:mode_saturation}
\scriptsize
\resizebox{\textwidth}{!}{%
\begin{ruledtabular}
\begin{tabular}{c cc cc cc cc}
&\multicolumn{2}{c}{$(c,+)$}&\multicolumn{2}{c}{$(c,-)$}&\multicolumn{2}{c}{$(b,+)$}&\multicolumn{2}{c}{$(b,-)$}\\
$N$&$\Delta E_N$&$\Cnf_N$&$\Delta E_N$&$\Cnf_N$&$\Delta E_N$&$\Cnf_N$&$\Delta E_N$&$\Cnf_N$\\
\hline
$1$&$0$&$0$&$0$&$0$&$0$&$0$&$0$&$0$\\
$2$&$0.663$&$0.00242$&$0.457$&$0.000370$&$0.658$&$0.00546$&$0.482$&$0.00125$\\
$3$&$0.858$&$0.02925$&$0.580$&$0.001024$&$0.863$&$0.03908$&$0.587$&$0.002790$\\
$5$&$1.201$&$0.02977$&$1.122$&$0.002349$&$1.254$&$0.04389$&$1.159$&$0.005909$\\
$10$&$1.861$&$0.06227$&$1.610$&$0.003233$&$1.800$&$0.1000$&$1.608$&$0.006198$\\
$20$&$2.555$&$0.09033$&$2.422$&$0.004990$&$2.531$&$0.1749$&$2.447$&$0.01172$\\
$50$&$4.493$&$0.1444$&$4.937$&$0.008338$&$4.384$&$0.2206$&$4.941$&$0.01655$\\
$100$&$9.504$&$0.2051$&$16.031$&$0.009112$&$9.218$&$0.3189$&$14.320$&$0.01817$\\
all&--&$0.2237$&--&$0.009324$&--&$0.3531$&--&$0.01854$
\end{tabular}
\end{ruledtabular}}
\end{table*}

A single nonzero mode gives rank one and a vanishing smaller eigenvalue.
The first three modes give only $11$--$15\%$ of the all-mode coefficient.
Across the four sectors, the excitation energy of the highest included pseudostate above the lowest mode is $4.4$--$4.9~\mathrm{GeV}$ for $N=50$ and $9$--$16~\mathrm{GeV}$ for $N=100$.
These high pseudostates serve as Gaussian basis components for the regulated short-distance sum, rather than as individual hybrid-resonance predictions.
The result is a broad short-distance sum in the regulated model.

An excitation-energy cut avoids comparing an equal number of modes from different discretizations.
We define $\Cnf_\eta^Q(E_{\max})$ by retaining modes with $E_n-E_1<E_{\max}$, keeping the same reference energy in the propagation denominators.

\begin{table*}[h]
\caption{Energy-cut coefficients. $N$ is the number of modes below the cut, and the parenthetical number is the fraction of the all-mode coefficient.}
\label{tab:s_energy_cut}
\label{tab:energy_cut_saturation}
\scriptsize
\begin{ruledtabular}
\begin{tabular}{c cc cc cc cc}
&\multicolumn{2}{c}{$(c,+)$}&\multicolumn{2}{c}{$(c,-)$}&\multicolumn{2}{c}{$(b,+)$}&\multicolumn{2}{c}{$(b,-)$}\\
$E_{\max}$ [GeV]&$N$&$\Cnf_\eta^Q(E_{\max})$&$N$&$\Cnf_\eta^Q(E_{\max})$&$N$&$\Cnf_\eta^Q(E_{\max})$&$N$&$\Cnf_\eta^Q(E_{\max})$\\
\hline
$0.5$&$1$&$0\ (0\%)$&$2$&$0.000370\ (4.0\%)$&$1$&$0\ (0\%)$&$2$&$0.001251\ (6.7\%)$\\
$1.0$&$3$&$0.02925\ (13.1\%)$&$4$&$0.002351\ (25.2\%)$&$3$&$0.03908\ (11.1\%)$&$4$&$0.005605\ (30.2\%)$\\
$2.0$&$11$&$0.06228\ (27.8\%)$&$14$&$0.003198\ (34.3\%)$&$13$&$0.11372\ (32.2\%)$&$13$&$0.007248\ (39.1\%)$
\end{tabular}
\end{ruledtabular}
\end{table*}

Mode-count and energy-cut results agree if they include the same intermediate space.
The energy-cut coefficients are listed in Table~\ref{tab:s_energy_cut}.
An energy cut compares spectral contributions over the same excitation interval, even when the two bases have different level densities.

\section{Relative offsets of the exchange sectors}

\begin{figure}[h]
\includegraphics[width=0.75\textwidth]{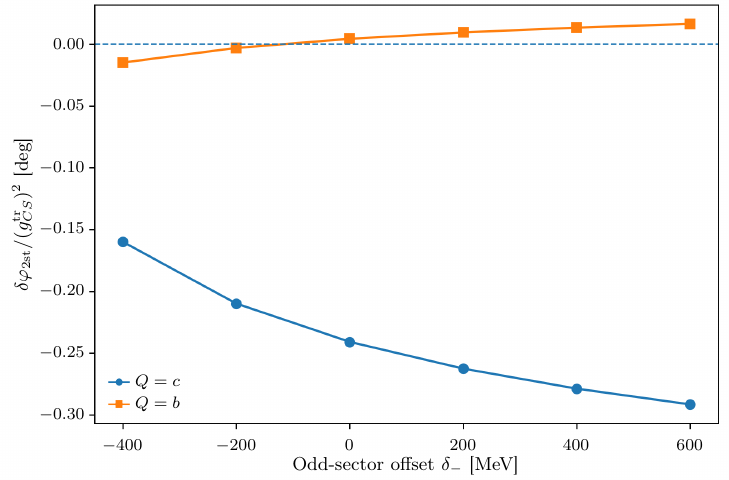}
\caption{Dependence of the linear \mbox{spin-$3/2$} retained-eigenbasis rotation $\delta\varphi_{\rm 2st}/(\gtr)^2$ in Eq.~\eqref{eq:s_signed_rotation} on the relative exchange-sector offset.}
\label{fig:s_sector_offset}
\end{figure}

This section isolates the set A $j_{Qg}=3/2$ contribution.
We combine the 270 even and 210 odd directions selected at $x=0$.
The lowest even level is placed at the reference gap $870~\MeV$ and the odd level at $870+\delta_-~\MeV$ before evaluating the flavor-breaking Hamiltonian.
Table~\ref{tab:s_offset} gives the Gram matrix in the valence phase convention of the main text.
Its off-diagonal sign changes if one retained state is rephased.
The leading signed rotation of the two-state eigenbasis is
\begin{equation}
 \delta\varphi_{\rm 2st}=\frac{(\gtr)^2G_{\rho\lambda}}{E_\lambda-E_\rho}.
 \label{eq:s_signed_rotation}
\end{equation}
Figure~\ref{fig:s_sector_offset} shows this linear rotation per unit squared coupling.
Table~\ref{tab:s_angle_direction} gives the norm angles obtained by diagonalizing $H_{\mathsf P\mathsf P}-(\gtr)^2G$ and evaluating $\theta_N=\tan^{-1}\sqrt{P_{\rm sub}/P_{\rm dom}}$ from the resulting wave function.

\begin{table*}[h]
\caption{Offset scan in the set A \mbox{spin-$3/2$} calculation at the fitted masses.
Gram entries are in MeV with the common coupling removed.
The last column gives the linear eigenbasis rotation per $(\gtr)^2$; probability-angle changes are evaluated separately from the wave functions.}
\label{tab:s_offset}
\begin{ruledtabular}
\begin{tabular}{c r r r r r}
$Q$&$\delta_-~[\MeV]$&$G_{\rho\rho}~[\MeV]$&$G_{\rho\lambda}~[\MeV]$&$G_{\lambda\lambda}~[\MeV]$&$\delta\varphi_{\rm 2st}~[\mathrm{deg}]$\\
\hline
$c$&$-400$&$5.0208$&$-0.2674$&$3.0076$&$-0.1600$\\
$c$&$-200$&$4.2511$&$-0.3509$&$2.9977$&$-0.2100$\\
$c$&$0$&$3.7536$&$-0.4027$&$2.9918$&$-0.2410$\\
$c$&$200$&$3.3952$&$-0.4389$&$2.9876$&$-0.2626$\\
$c$&$400$&$3.1198$&$-0.4660$&$2.9846$&$-0.2788$\\
$c$&$600$&$2.8989$&$-0.4873$&$2.9822$&$-0.2916$\\
\hline
$b$&$-400$&$0.9740$&$-0.03323$&$0.5140$&$-0.01491$\\
$b$&$-200$&$0.8174$&$-0.00682$&$0.5094$&$-0.00306$\\
$b$&$0$&$0.7180$&$0.00959$&$0.5066$&$0.00430$\\
$b$&$200$&$0.6472$&$0.02108$&$0.5046$&$0.00946$\\
$b$&$400$&$0.5932$&$0.02974$&$0.5031$&$0.01334$\\
$b$&$600$&$0.5501$&$0.03658$&$0.5020$&$0.01641$
\end{tabular}
\end{ruledtabular}
\end{table*}

The charm norm angle decreases over the \mbox{spin-$3/2$} offset scan, whereas the minimal spin-complete calculation gives an upward shift.

\begin{table}[h]
\caption{Norm angle after diagonalizing the \mbox{spin-$3/2$}-corrected retained matrix.
The column $\theta_{N,0}$ is the valence-only value, and $\theta_N$ is the corrected value.
Each row lists the common strength $\gtr$.}
\label{tab:s_angle_direction}
\begin{ruledtabular}
\begin{tabular}{c r c c c}
$Q$&$\delta_-~[\MeV]$&$\gtr$&$\theta_{N,0}$&$\theta_N$\\
\hline
$c$&$-400$&$1.0$&$3.114^\circ$&$2.989^\circ$\\
$c$&$0$&$1.0$&$3.114^\circ$&$2.924^\circ$\\
$c$&$600$&$1.0$&$3.114^\circ$&$2.883^\circ$\\
$c$&$0$&$2.2$&$3.114^\circ$&$2.305^\circ$\\
$b$&$-400$&$1.0$&$0.895^\circ$&$0.903^\circ$\\
$b$&$0$&$1.0$&$0.895^\circ$&$0.892^\circ$\\
$b$&$600$&$1.0$&$0.895^\circ$&$0.886^\circ$
\end{tabular}
\end{ruledtabular}
\end{table}

\begin{samepage}
At equal offsets and unit coupling, the linear formula in Eq.~\eqref{eq:s_signed_rotation} gives $-0.24096^\circ$ for charm.
Exact diagonalization of the retained $2\times2$ matrix gives $-0.23905^\circ$, while the probability angle changes by $-0.18947^\circ$.
These are different definitions, since the two valence states contain different optimized spatial amplitudes.
The model splitting is $95.75~\MeV$, compared with the observed separation of approximately $108~\MeV$.
Using the latter denominator reduces the magnitude of the linear estimate by about $11\%$.
\par
\end{samepage}

\section{Expansion of the projected spin coefficients}

We use the effective two-state spin-matrix convention defined in the main text.
The three coefficients of the reduced Gram matrix have the exchange-symmetry expansions
\begin{align}
 \alpha_Q(x)&=\alpha_{0,Q}+\alpha_{2,Q}x^2+O(x^4),\nonumber\\
 \beta_Q(x)&=\beta_{0,Q}+\beta_{2,Q}x^2+O(x^4),\nonumber\\
 \chi_Q(x)&=\chi_{1,Q}x+O(x^3).
\end{align}
Fitting over $0\leq x\leq0.008$ gives
\begin{equation}
\begin{array}[b]{c|rrrrr}
Q&\alpha_0&\alpha_2&\beta_0&\beta_2&\chi_1\\\hline
c&3.18399&-0.07584&-0.22135&+0.32860&+0.37389\\
b&0.56525&-0.15367&-0.05776&+0.06180&-0.001137
\end{array}.
\label{eq:s_local_small_x}
\end{equation}
All entries are in MeV before multiplication by $-(\gtr)^2$, and $x$ is dimensionless.
In the redundant pair-spin representation, the convention $A_Q=0$ gives
\begin{equation}
\begin{array}[b]{c|rrrrr}
Q&B_0&B_2&C_{+,0}&C_{+,2}&C_{-,1}\\\hline
c&-1.28268&+0.35388&-1.06133&+0.02528&+0.37389\\
b&-0.24618&+0.11302&-0.18842&+0.05122&-0.001137
\end{array}.
\label{eq:s_pair_operator_small_x}
\end{equation}
These coefficients parametrize the retained matrix using the pure-spin matrices in Appendix~A of the main text.

\section{Nondegenerate reference energies}

The common-energy Feshbach expression uses $G=W^\dagger \mathbf D^{-1}W$ at one energy $E$.
For distinct unperturbed valence energies, the Hermitian second-order expression is
\begin{equation}
 (G_{\rm SW})_{\alpha\beta}=\frac12\sum_nW_{n\alpha}^*W_{n\beta}
 \left[\frac{1}{E_{h_n}-E_\alpha}+\frac{1}{E_{h_n}-E_\beta}\right].
 \label{eq:s_nondegenerate}
\end{equation}
The physical correction is $-(\gtr)^2G_{\rm SW}$.
Table~\ref{tab:s_sw} compares the two prescriptions for the set A \mbox{spin-$3/2$} calculation at equal offsets.

\begin{table}[h]
\caption{Common-energy Gram entries and nondegenerate second-order entries.
Both are given in MeV with the common coupling removed.
The last column gives the relative change from the common-energy value.}
\label{tab:s_sw}
\begin{ruledtabular}
\begin{tabular}{c c r r r}
$Q$&entry&common&nondegenerate&relative change\\
\hline
$c$&$G_{\rho\rho}$&$3.7536$&$3.6533$&$-2.67\%$\\
$c$&$G_{\rho\lambda}$&$-0.40269$&$-0.40253$&$-0.04\%$\\
$c$&$G_{\lambda\lambda}$&$2.9918$&$3.1115$&$+4.00\%$\\
$b$&$G_{\rho\rho}$&$0.71798$&$0.69183$&$-3.64\%$\\
$b$&$G_{\rho\lambda}$&$0.009585$&$0.009373$&$-2.21\%$\\
$b$&$G_{\lambda\lambda}$&$0.50657$&$0.53496$&$+5.60\%$
\end{tabular}
\end{ruledtabular}
\end{table}

The valence splittings are $95.75~\MeV$ for charm and $127.69~\MeV$ for bottom.
The diagonal entries change by a few percent in this comparison.
The symmetrized expression is Hermitian, but positivity is not guaranteed for general transitions.
The exact single-direction criterion applies to the positive common-energy Feshbach expression.

\section{Reference normalization and spin-channel dependence}

We compare two prescriptions for setting $\gtr$ from the reference mass shifts.
The first bounds each diagonal hybrid-shift magnitude by the corresponding reference value, $12.123$ or $14.080~\MeV$.
As discussed in Sec.~VI of the main text, the reference interaction and full hybrid self-energy have different lower-body subtractions.
For the set A \mbox{spin-$3/2$} matrix at equal offsets, the condition gives
\begin{equation}
 (g_{CS}^{\rm tr})^2\leq3.2297,
 \qquad g_{CS}^{\rm tr}\leq1.7971,
 \qquad \theta_N=2.5412^\circ,
\end{equation}
where the quoted angle is evaluated at the upper allowed strength.
The unrestricted norm-angle scan has a minimum $1.83661^\circ$ at $(\gtr)^2=11.4215$.
Set B has a corresponding minimum $1.412^\circ$.

An alternative prescription fits both reference magnitudes in least squares with one strength.
For spin $3/2$ alone, it gives
\begin{equation}
 (g_{CS}^{\rm tr})^2_{\rm sat}=3.8033,
 \qquad g_{CS}^{\rm tr}=1.9502,
 \qquad \theta_N=2.4528^\circ,
\end{equation}
and predicted diagonal magnitudes $(14.28,11.38)~\MeV$, rather than $(12.123,14.080)~\MeV$.
This least-squares prescription minimizes the combined residual independently of the separate upper-bound condition.
Table~\ref{tab:s_parent_shape} compares the one-strength results with the reference shifts and quantifies the remaining mismatch in the two diagonal entries.

\begin{table*}[h]
\footnotesize
\caption{One-strength comparison of the two reference diagonal shifts.
The reference magnitudes are $(12.123,14.080)~\MeV$.
The relative residual is the Euclidean residual norm divided by the norm of this reference vector.
}
\label{tab:s_parent_shape}
\begin{ruledtabular}
\setlength{\tabcolsep}{4pt}
\begin{tabular}{@{}l c c c c@{}}
resolved topology&\shortstack{best\\$(g_{CS}^{\rm tr})^2$}&\shortstack{predicted $(|\Delta E_\rho|,|\Delta E_\lambda|)$\\(MeV)}&\shortstack{diagonal\\ratio}&\shortstack{relative\\residual}\\
\hline
$j_{Qg}=3/2$&$3.8033$&$(14.28,11.38)$&$1.255$&$18.6\%$\\
minimal spin complete&$1.2679$&$(8.24,16.07)$&$0.513$&$23.5\%$\\
parent target&--&$(12.123,14.080)$&$0.861$&--
\end{tabular}
\end{ruledtabular}
\end{table*}

The direction of the \mbox{spin-$3/2$} correction over the offset scan is different from that of the spin-complete result.
Table~\ref{tab:s_sign_completion} gives a unit-coupling benchmark of the set A \mbox{spin-$3/2$} model in Sec. X, with $\zeta_{1/2}=0$.
For each listed offset, we diagonalize $H_{\mathsf P\mathsf P}-G$ using the Gram entries of Table~\ref{tab:s_offset} and evaluate the spin probabilities of the resulting superposition of the two valence eigenstates.
The valence wave functions use the same set A seven-range basis as the spectrum check in Sec.~\ref{sec:s_valence_spectrum}.


\begin{table*}[h]
\caption{Set A \mbox{spin-$3/2$} unit-coupling norm-angle shifts, $\Delta\theta_{N,Q}=\theta_{N,Q}-\theta_{N,0,Q}$, at the displayed offsets.
The input Gram matrices are those of the 480-direction calculation in Table~\ref{tab:s_offset}, with $\gtr=1$ and $\zeta_{1/2}=0$.
The shifts are evaluated before rounding the angles and reproduce the selected rounded values in Table~\ref{tab:s_angle_direction}.
The norm-angle changes include the spin probabilities of the mixed wave functions, which distinguish them from the eigenbasis rotation in Eq.~\eqref{eq:s_signed_rotation}.}
\label{tab:s_sign_completion}
\begin{ruledtabular}
\begin{tabular}{r r r}
$\delta_-\,[\MeV]$&$\Delta\theta_{N,c}\,[\mathrm{deg}]$&$\Delta\theta_{N,b}\,[\mathrm{deg}]$\\
\hline
$-400$&$-0.1250$&$+0.0078$\\
$-200$&$-0.1647$&$+0.0016$\\
$0$&$-0.1895$&$-0.0022$\\
$200$&$-0.2069$&$-0.0049$\\
$400$&$-0.2200$&$-0.0069$\\
$600$&$-0.2303$&$-0.0084$
\end{tabular}
\end{ruledtabular}
\end{table*}

For the minimal vector vertex, $\zeta_{1/2}=1$.
The spin sectors propagate independently, and their relative phase cancels in the mass correction.
Changing $\zeta_{1/2}$ tests additional transition structures, while changing their gaps tests the diagonal Hamiltonian.
The table benchmarks the $\zeta_{1/2}=0$ limit, while the results below use the separately normalized minimal vertex at $\zeta_{1/2}=1$.

Table~\ref{tab:s_spin_complete_budget} gives the minimal spin-complete result with least-squares reference normalization.
The hybrid probability is obtained from the reconstructed norm $(\gtr)^2p^\dagger W^\dagger \mathbf D^{-2}Wp$ using the normalization denominator in Eq.~\eqref{eq:s_hybrid_probability}.

\begin{table*}[h]
\caption{Minimal spin-complete ranges with one-strength reference normalization.
The scans use $670\leq\Delta_{1/2}\leq1070~\MeV$ and $-400\leq\delta_-\leq600~\MeV$.
The charm-determined coupling is also used for bottom.
The listed angles are norm angles; Table~\ref{tab:s_angle_definitions} also gives the source and mass-operator values.}
\label{tab:s_spin_complete_budget}
\begin{ruledtabular}
\begin{tabular}{c c c c c}
quantity&minimum&maximum&quantity&range\\
\hline
$(g_{CS}^{\rm tr})^2$&$1.037$&$1.430$&$\theta_{N,c}$&$3.268$--$3.662^\circ$\\
$P_{\rm hyb}^c$&$0.279\%$&$1.606\%$&$\theta_{N,b}$&$0.892$--$0.908^\circ$\\
$P_{\rm hyb}^b$&$0.045\%$&$0.288\%$&$|\Delta\theta_{N,b}|$&$<0.014^\circ$
\end{tabular}
\end{ruledtabular}
\end{table*}

The wider conditional upper-bound scan over $0\leq\zeta_{1/2}\leq2$ gives
\[
 -1.248^\circ\leq\Delta\theta_{N,c}\leq+1.028^\circ,
 \qquad P_{\rm hyb}<1.68\%.
\]
The valence norm angle in set B is $2.808^\circ$, and its unrestricted \mbox{spin-$3/2$} minimum is $1.412^\circ$ at $(\gtr)^2=5.58$.

\section{Effective currents and hybrid normalization}

Let $p$ be a retained vector with $p^\dagger p=1$.
The leading reconstructed hybrid component is
\begin{equation}
 |\psi_{\mathsf Q}\rangle=-\gtr \mathbf D^{-1}Wp.
\end{equation}
The minus sign follows from $E-H_{\mathsf Q\mathsf Q}=-\mathbf D$ in the hybrid energy basis.
Define $\Omega_Q=W^\dagger \mathbf D^{-2}W$ and $\omega_p=p^\dagger \Omega_Qp$.
Normalizing the combined retained and hybrid state gives
\begin{equation}
 P_{\rm hyb}=\frac{(\gtr)^2\omega_p}{1+(\gtr)^2\omega_p}.
 \label{eq:s_hybrid_probability}
\end{equation}

For an external current $\mathcal J$, write $\mathcal J_{\mathsf A\mathsf B}=\mathsf A\mathcal J\mathsf B$, where $\mathsf A,\mathsf B\in\{\mathsf P,\mathsf Q\}$ are the projectors onto the output and input spaces, respectively.
The normalized effective current at a common reference energy is
\begin{align}
 \mathcal J_{\rm eff}={}&\mathcal J_{\mathsf P\mathsf P}-\gtr \mathcal J_{\mathsf P\mathsf Q}\mathbf D^{-1}W
 -\gtr W^\dagger \mathbf D^{-1}\mathcal J_{\mathsf Q\mathsf P}\nonumber\\
 &+(\gtr)^2W^\dagger \mathbf D^{-1}\mathcal J_{\mathsf Q\mathsf Q}\mathbf D^{-1}W\nonumber\\
 &-\frac{(\gtr)^2}{2}\{\Omega_Q,\mathcal J_{\mathsf P\mathsf P}\}+O((\gtr)^3).
 \label{eq:s_effective_current}
\end{align}
The anticommutator term is the normalization correction.
If the current conserves explicit gluon number, $\mathcal J_{\mathsf P\mathsf Q}=\mathcal J_{\mathsf Q\mathsf P}=0$ and the leading correction is quadratic in the transition amplitude.
The hybrid probability then sets the wave-function normalization scale, while each current matrix element determines its own relative correction.
A small leading decay amplitude can have a large relative correction even if this scale is small.
Transition currents can enter linearly, with a wave-function scale of order $\sqrt{P_{\rm hyb}}$ multiplied by the reduced current matrix elements.

For a transition between different sectors, each state has its own resolvent and normalization.
Let $B_i=-g_iD_i^{-1}W_i$ and $B_f=-g_fD_f^{-1}W_f$ be their reconstructed hybrid maps, and let $N_a=I+B_a^\dagger B_a$ for $a=i,f$.
The normalized current is
\begin{equation}
 \begin{aligned}
 \mathcal J_{{\rm eff},fi}=N_f^{-1/2}\bigl[&\mathcal J_{\mathsf P\mathsf P}^{fi}+\mathcal J_{\mathsf P\mathsf Q}^{fi}B_i
 +B_f^\dagger \mathcal J_{\mathsf Q\mathsf P}^{fi}\\
 &+B_f^\dagger \mathcal J_{\mathsf Q\mathsf Q}^{fi}B_i\bigr]N_i^{-1/2}.
 \end{aligned}
 \label{eq:s_two_energy_current}
\end{equation}
The superscript $fi$ identifies the final and initial model-space blocks.
The strengths $g_i,g_f$ may be set equal if the model assumes a common coupling.
Expanding the normalization factors to second order gives the corresponding transition-current formula.
Equation~\eqref{eq:s_effective_current} is the common-sector, common-reference limit.
The resulting current matrix elements determine the radiative and semileptonic amplitudes.

\Needspace{8\baselineskip}
\section{Radial form of the transition}
\label{sec:s_transition_shape}

We compare the Gaussian radial profile $u_{iQ}(r)$ of Eq.~(42) in the main text with a Yukawa profile having the same three-dimensional integral and root-mean-square (rms) radius.
The latter is defined by the profile-weighted moment
\[
 r_{{\rm rms},u}^2
 =\frac{\int d^3r\,r^2u_{iQ}(r)}{\int d^3r\,u_{iQ}(r)}
 =\left(\frac{\hbar c}{\mu_{iQ}}\right)^2.
\]
The weight is the radial transition profile $u_{iQ}$ itself.
For each flavor pair, the replacement is
\[
 \frac{e^{-(\mu r/\hbar c)^2}}{r}
 \longrightarrow \frac{3e^{-\sqrt6\,\mu r/(\hbar c)}}{r}.
\]
Here $\mu=\mu_{iQ}$ is the range scale for the corresponding flavor pair.
For the two displayed radial factors, the three-dimensional integrals both equal $2\pi(\hbar c/\mu)^2$, and the profile-weighted mean-square radii both equal $(\hbar c/\mu)^2$.
Only this radial factor is changed; its prefactor and the separate $Qg$ Gaussian are kept fixed.

\begin{table*}[h]
\caption{Nonfactorization coefficients for the Gaussian and equal-integral, equal-rms-radius Yukawa transitions.
The last column is the Yukawa-to-Gaussian ratio.}
\label{tab:s_transition_shape}
\begin{ruledtabular}
\begin{tabular}{c c c c c}
$Q$&$\eta$&$\Cnf_{\rm G}$&$\Cnf_{\rm Y}$&$\Cnf_{\rm Y}/\Cnf_{\rm G}$\\
\hline
$c$&$+$&$0.2237$&$0.2546$&$1.138$\\
$c$&$-$&$0.009324$&$0.01092$&$1.171$\\
$b$&$+$&$0.3531$&$0.4010$&$1.136$\\
$b$&$-$&$0.01854$&$0.02017$&$1.088$
\end{tabular}
\end{ruledtabular}
\end{table*}

Table~\ref{tab:s_transition_shape} shows that the coefficients change by $8.8$--$17.1\%$ and remain nonzero in all four sectors.

\Needspace{10\baselineskip}
\section{Interference of the two responses}

Let $u_{\rm vert}$ and $u_{\rm resp}$ be the perpendicular vertex and state-response vectors; their sum is the full response.
With $\cos\phi$ denoting their normalized real inner product,
\begin{equation}
 \Cnf_{\rm full}=\Cnf_{\rm vert}+\Cnf_{\rm resp}
 +2\cos\phi\sqrt{\Cnf_{\rm vert}\Cnf_{\rm resp}}.
\end{equation}
The extracted values are
\begin{equation}
\begin{array}[b]{c|rrrr}
(Q,\eta)&(c,+)&(c,-)&(b,+)&(b,-)\\\hline
\cos\phi&+0.234&-0.673&+0.307&-0.636
\end{array}.
\end{equation}
Positive values indicate constructive interference in the even sectors, and negative values indicate destructive interference in the odd sectors.
The cross term explicitly accounts for interference between the two response mechanisms.

\section{Basis and Hamiltonian sensitivity}

\begin{figure}[h]
\includegraphics[width=0.82\textwidth]{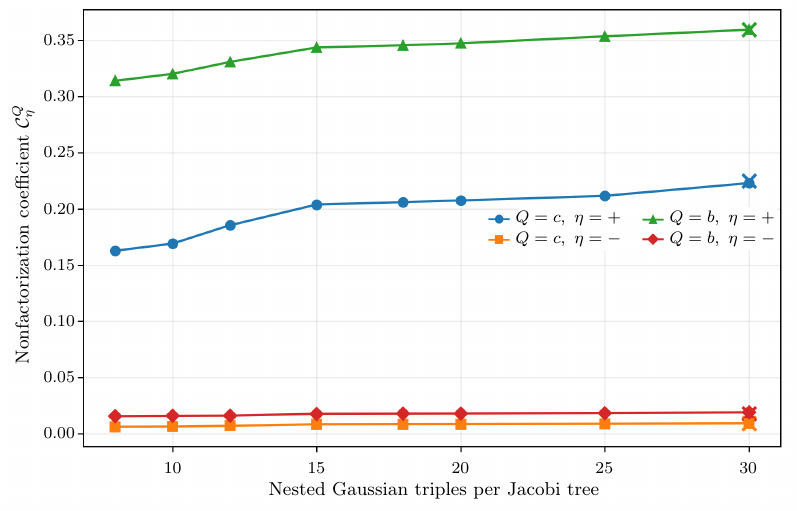}
\caption{Nonfactorization coefficients in nested per-tree bases.
The crosses at 30 triples show the independent geometric-log calculation.}
\label{fig:s_nested_basis}
\end{figure}

Allocating the ranges for all Jacobi trees from one Sobol sequence gives
\begin{equation}
\begin{array}[b]{c|rrrrrr}
N_{\rm triple}/\text{tree}&8&10&12&15&18&20\\\hline
\gamma_-~[\keV]&39.18&36.85&37.82&45.13&46.01&45.58
\end{array}.
\label{eq:s_old_nonnested_sequence}
\end{equation}
In this construction, changing the number of triples changes the allocation of ranges to each tree, so the bases are not nested.
For a nested comparison, we generate an independent scrambled sequence for each tree and retain its prefixes as the basis grows.
Table~\ref{tab:s_nested_basis} and Fig.~\ref{fig:s_nested_basis} compare the four nonfactorization coefficients through 30 triples per tree with a deterministic geometric-log construction.

\begin{table*}[h]
\caption{Nested-basis convergence of $\Cnf_\eta^Q$. ``Geom. 30'' uses a deterministic geometric-log range construction rather than Sobol points.}
\label{tab:s_nested_basis}
\begin{ruledtabular}
\begin{tabular}{c c r r r r r}
$Q$&$\eta$&nested 20&nested 25&nested 30&geom. 30&15-triple value\\
\hline
$c$&$+$&$0.20754$&$0.21167$&$0.22295$&$0.22508$&$0.22374$\\
$c$&$-$&$0.008567$&$0.008818$&$0.009244$&$0.009082$&$0.009324$\\
$b$&$+$&$0.34733$&$0.35357$&$0.35943$&$0.35966$&$0.35310$\\
$b$&$-$&$0.017982$&$0.018383$&$0.018942$&$0.017991$&$0.018543$
\end{tabular}
\end{ruledtabular}
\end{table*}

The nested $25\to30$ changes are $1.7$--$5.3\%$.
The geometric-log and nested 30-triple results differ by $0.1$--$5.0\%$, while the 15-triple values differ from nested 30 by $0.4$--$2.2\%$.
We use a conservative $6\%$ numerical uncertainty for the integrated coefficients; convergence of individual high pseudostate energies requires separate tests.

\begin{figure}[h]
\includegraphics[width=0.75\textwidth]{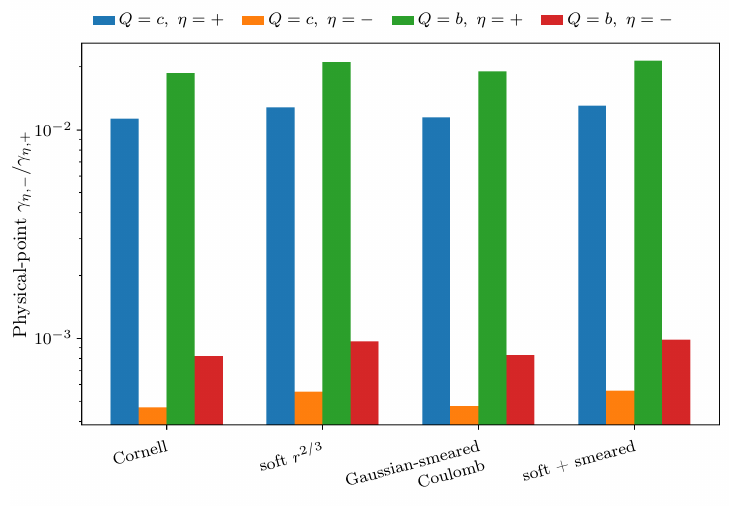}
\caption{Smaller-to-larger Gram eigenvalue ratios at the fitted mass point for alternative central hybrid interactions.}
\label{fig:s_dynamics}
\end{figure}

We also compare Cornell, $r^{2/3}$ confinement, Gaussian-smeared Coulomb, and combined soft-confinement/smeared-Coulomb hybrid Hamiltonians in a common 10-triple basis.
Their fitted small-$x$ powers range from $1.99938$ to $2.00011$.
The near-quadratic behavior is required by exchange symmetry for a nonzero leading coefficient.
All four hybrid Hamiltonians give nonzero nonfactorization coefficients.
Figure~\ref{fig:s_dynamics} shows the change of the eigenvalue ratios.
Table~\ref{tab:s_regulator_stability} gives two further tests of the $Qg$ Gaussian width and the reference gap.

\begin{table*}[h]
\caption{Relative changes under scans of the $Qg$ Gaussian width $r_g$ and the common reference gap.
Each entry is $\Cnf_\eta^Q/\Cnf_{\eta,\mathrm{baseline}}^Q$.
Both exchange sectors use the common reference gap $\Delta_0$.}
\label{tab:s_regulator_stability}
\begin{ruledtabular}
\begin{tabular}{l c c c c c}
variation&value&$(c,+)$&$(c,-)$&$(b,+)$&$(b,-)$\\
\hline
$r_g$ [fm]&$0.25$&$1.014$&$1.017$&$1.011$&$1.016$\\
&$0.30$&$1.000$&$1.000$&$1.000$&$1.000$\\
&$0.40$&$0.977$&$0.943$&$0.995$&$0.962$\\
\hline
$\Delta_0$ [MeV]&$700$&$0.904$&$0.933$&$0.903$&$0.934$\\
&$870$&$1.000$&$1.000$&$1.000$&$1.000$\\
&$1050$&$1.084$&$1.057$&$1.085$&$1.056$
\end{tabular}
\end{ruledtabular}
\end{table*}

The maximum changes are $5.7\%$ for the $Qg$ Gaussian width $r_g$ and $9.7\%$ for the reference gap.

The two valence parameter sets give
\begin{equation}
 \Knf_+^c=0.6628\ \text{and}\ 0.2651~\MeV,
\end{equation}
while their fitted mass coordinates are $x=0.2851$ and $0.3057$.
Since both coefficients are extracted near the symmetric point, their difference measures valence-model dependence separately from the physical mass coordinate.
The valence-model dependence is appreciable and should be kept separate from numerical basis uncertainty.

\section{Check using the finite detuning matrix}

For an implementation check, introduce a real detuning parameter $\tau$ and the finite matrix
\begin{equation}
 H(\tau)=
 \begin{pmatrix}
 \tau I_2&\gtr W^\dagger\\
 \gtr W&\mathbf D-\tau I
 \end{pmatrix}.
\end{equation}
The identities $I_2$ and $I$ act in the retained and intermediate spaces, respectively.
We vary $\tau$ until the eigenvalue connected to the smaller Gram direction crosses zero.
For $\mathbf D-\tau I$ invertible, the zero-eigenvalue condition is
\[
 \det\!\left[\tau I_2-(\gtr)^2W^\dagger(\mathbf D-\tau I)^{-1}W\right]=0.
\]
The small positive root approaches $\tau=(\gtr)^2\gamma_-$ if $|\tau|/D_{\min}$ is small.
At $\gtr=1$, the reported maximum relative difference is $1.4\times10^{-5}$.
This comparison tests the reduction and the root-finding procedure.